\documentclass[prl,twocolumn,10pt,%
nobibnotes,   
 amsmath,amssymb,
 aps,
]{revtex4-2}

\pdfoutput=1
\usepackage{graphicx}
\usepackage{dcolumn}
\usepackage{bm}
\usepackage[caption=false]{subfig}
\usepackage{amssymb}
\usepackage{amsmath}
\usepackage{commath}
\usepackage{graphicx,bm}
\usepackage{verbatim}

\usepackage[colorlinks=true,linkcolor=blue,citecolor=blue,allcolors=blue]{hyperref}%

\usepackage{color}
    \definecolor{darkgreen}{rgb}{0,0.5,0}
    \definecolor{darkred}{rgb}{0.5,0,0}
    \definecolor{darkblue}{rgb}{0,0,0.6}
    \definecolor{purple}{rgb}{0.4,.2,0.7}

\def \dd {\mathrm{d}}

\begin{document}

\preprint{APS/123-QED}

\title{Spinning Black Binaries in de Sitter space}

\author{\'Oscar~J.~C.~Dias}
\email{ojcd1r13@soton.ac.uk}
\affiliation{STAG research centre \& Mathematical Sciences, University of Southampton, Highfield Campus, UK}
\author{Jorge~E.~Santos}
\email{jss55@cam.ac.uk}
\affiliation{DAMTP, Centre for Mathematical Sciences, University of Cambridge, Wilberforce Road, Cambridge CB3 0WA, UK}%
\author{Benson~Way}
\email{bway@umd.edu}
\affiliation{Maryland Center for Fundamental Physics, University of Maryland, College Park, MD 20742, USA}

\begin{abstract}
We construct stationary, rotating black binaries in general relativity with a positive cosmological constant.  We consider identical black holes with either aligned or anti-aligned spins.  Both cases have less entropy than the corresponding single Kerr/Schwarzschild de Sitter black hole with the same total angular momentum and cosmological horizon entropy.  Our solutions establish continuous non-uniqueness in general relativity without matter.  They also provide initial data for the spinning binary merger problem (when orbital angular momentum is added).
\end{abstract}

\maketitle
\subparagraph{Introduction.}

A single black hole in isolation is an unchanging, time-independent gravitational object.  But can two or more black holes exist in such a state of stationary equilibrium?  This question has been around since the early days of general relativity.  Early work by Bach and Weyl \cite{BachWeyl(Republication):1922}, later generalized by Israel and Khan \cite{Israel1964} found multi-black hole solutions that were static, but contained unphysical conical singularities.  Indeed, several theorems \cite{Israel:1967wq,1977GReGr...8..695R,1987GReGr..19..147B} preclude the existence of static, asymptotically flat multi black holes.  With the addition of electric charge, a well-known multi-black hole solution was found by Majumdar and Papapetrou \cite{Majumdar:1947eu,Papapetrou:1947ib}, but these solutions require that the black holes be maximally charged, which is unrealistic.  Multi black holes were also found in more exotic situations, such as higher-dimensional spacetimes with compact directions \cite{Astorino:2022fge}.

A natural consideration that was not included above is the effect of rotation.  While orbital rotation breaks stationarity due to the emission of gravitational waves, spin-spin interactions \cite{Wald:1972sz} do not.  Notably, spin-spin interactions are a key ingredient to molecular stability \cite{Bohr1913a,Bohr1913b,Bohr1913c,Svidzinsky2014}. 
Newtonian attraction scales with distance as $1/r^2$, while spin-spin interaction scales as $1/r^3$ and therefore might have a dominant role at smaller distances. Nevertheless, all known asymptotically flat, stationary black binaries are singular, even those that maximize spin-spin repulsion (see \emph{e.g.} \cite{PerjesPhysRevLett.27.1668,Israel:1972vx,Hartle:1972ya,Tomimatsu:1972zz,KRAMER1980259,NEUGEBAUER198191,stephani_kramer_maccallum_hoenselaers_herlt_2003,Manko2001,Herdeiro:2008kq,Manko:2008pv,Manko:2017avt,Manko:2018iyn,Manko:2020jfa}), though the existence of singularity-free spinning binaries has not been ruled out by any theorem.  We do mention here that with the addition of a massive complex scalar field, the scalar pressure together with the spin-spin can balance a spinning binary of hairy black holes \cite{Herdeiro:2023roz}.

What about the inclusion of a positive cosmological constant ($\Lambda>0$)?  After all, this ingredient explains the accelerated expansion of the Universe \cite{SupernovaCosmologyProject:1997zqe,SupernovaSearchTeam:1998fmf,SupernovaCosmologyProject:1998vns,Wright2011ER,Planck:2013pxb}, at least according to the $\Lambda$-Cold Dark Matter ($\Lambda$CDM) model.  With a positive cosmological constant, it was recently found that non-rotating black binary solutions do indeed exist \cite{Dias:2023rde}.  These solutions are called de Sitter black binaries.  These binaries exist because the expansion from the cosmological constant balances out the mutual attraction of the black holes, and their existence could be anticipated from Newton-Hooke analysis \cite{Dias:2023rde}.  However, this configuration is unstable, as any perturbation would cause the black holes to merge or fly apart.

In this Letter, we add rotation to the de Sitter binaries found in \cite{Dias:2023rde}, study their properties, and discuss the prospect that spin-spin interactions can stabilize the binaries.  For simplicity, we take the individual black holes to be identical and focus on the cases where the spin axes are aligned or anti-aligned, which happens to also be the situations with the strongest spin-spin interaction.

Our results are also of mathematical interest, as they establish continuous non-uniqueness of de Sitter black holes, in contrast to the asymptotically flat case \cite{Chase:1970,Penney:1968zz,Bekenstein:1972ny,Bekenstein:1971hc,Bekenstein:1972ky,Teitelboim:1972qx,Hartle1972,Heusler:1992ss,Bekenstein:1995un,Bekenstein:1996pn,Sudarsky:1995zg}.

\subparagraph{Newton-Hooke.}
Some de Sitter binaries can be described by Newton-Hooke analysis \cite{Dias:2023rde}, which we review here with the inclusion of spin-spin interactions.  We adopt geometrized units $c=G_N=k_B=\hbar=1$.

Consider $N$ black holes with masses $m_a$, positions $\mathbf x_a$, and spin vectors $\mathcal{S}_a$, $a=1,\ldots ,N$.  Including a gravitational force, cosmological Hooke's law, and dipolar spin-spin interactions, Newton's second law becomess
\begin{equation}\label{NewtonHooke}
\begin{aligned}
m_a \frac{\mathrm{d}^2\mathbf{x}_a}{\mathrm{d}t^2}
=& \,m_a \frac{\mathbf{x}_a}{\ell^2}+ \nabla \Big(\frac{\,m_a\,m_b}{|\mathbf{r}_{ab}|} \Big) \\
 & + \nabla \Bigg[\frac{\mathcal{S}_a\cdot \mathcal{S}_b}{|\mathbf{r}_{ab}|^3}
  -\frac{3 \left(\mathcal{S}_a\cdot\mathbf{r}_{ab} \right) \left(\mathcal{S}_b\cdot\mathbf{r}_{ab} \right)}{|\mathbf{r}_{ab}|^5} \Bigg]\,,
  \end{aligned}
\end{equation}
for $b\neq a$, de Sitter radius $\ell=\sqrt{\Lambda/3}$, and $\mathbf{r}_{ab}\equiv \mathbf{x}_a-\mathbf{x}_b$.  Stationary solutions exist when $\frac{\mathrm{d}^2\mathbf{x}_a}{\mathrm{d}t^2}=0$.

Consider now the case with two equal mass black holes separated by a distance $d$, aligned along the $z$-axis, with their spins also aligned along the $z$-axis: $m_1=m_2\equiv m$, $\mathbf{x}_1=-\mathbf{x_2}=\frac{z}{2}\,\mathbf{e}_{z}$ and $\mathcal{S}_{1,2}=m \,\sigma_{1,2}\,\mathbf{e}_{z}$. Take the symmetric cases $\sigma_2=  \gamma \sigma_1 \equiv   \gamma \sigma $ with $\gamma=1$ (repulsive spin-spin force) or $\gamma=-1$ (attractive spin-spin force).  Then \eqref{NewtonHooke} admits stationary spinning binary solutions when $\frac{d^3}{\ell^3}= \frac{2 m}{\ell} \left(1- 6\,\gamma \, \frac{\sigma^2}{d^2} \right)$, which implies
\begin{equation}\label{eq:NHpred}
\frac{d}{\ell} \simeq \frac{1}{\left(4\pi T_+ \ell\right)^{\frac{1}{3}}}\left\{1-\left[\frac{1}{3}+\frac{2\gamma}{\left(4\pi T_+ \ell\right)^{4/3}}\right]\frac{\Omega_+^2}{\left(4\pi T_+\right)^2}\right\}\,
\end{equation}
In deriving the above, we discarded $\mathcal{O}(\Omega_+^4)$ terms.  We also expressed the mass and spin of the test body in terms of the temperature $T_+$ and angular velocity $\Omega_+$ of a Kerr black hole, so that
\begin{equation}
\begin{aligned}
& 2\,m=\frac{1}{2\pi T_+ +\sqrt{4\pi^2T_+^2+\Omega_+^2}}\,,
\\
& 
\sigma=\frac{m\,\Omega_+}{\sqrt{4\pi^2T_+^2+\Omega_+^2}}\,.
 \end{aligned}
\end{equation}For vanishing spin, $\Omega_+=0$, (\ref{eq:NHpred}) yields $\frac{d^3}{\ell^3}= (4\pi T_+\ell )^{-\frac{1}{3}} = \frac{r_S}{\ell}$, where $r_S=2m$ is the Schwarzschild radius of the particles, in agreement with \cite{Dias:2023rde}.  The equilibrium condition (\ref{eq:NHpred}) can fall within the regime of validity of Newton-Hook theory, which occurs when $r_S\ll d\ll \ell$ (i.e. large $T_+\ell$). 

 With nonzero spin, $\Omega_+\neq  0$, an inspection of \eqref{eq:NHpred} for a given (large) $T_+\ell$ shows that $d/\ell$ is a monotonically decreasing function of $\Omega_+ \ell$. Moreover, the $d(\Omega_+)$ curve for the aligned binary ($\gamma=1$) is always below the curve for the anti-aligned case ($\gamma=-1$).
This agrees with expectations as spin-spin forces are repulsive for aligned spins, so the black holes need to be closer apart to remain in equilibrium (for fixed gravitational and cosmological forces).  The opposite is true for anti-aligned spins. Our numerical (anti-)aligned spinning solutions of de Sitter general relativity will match this behaviour.

\subparagraph{Numerical construction.}
Now we construct spinning de Sitter binaries using numerical relativity.  We use the Einstein-DeTurck method \cite{Headrick:2009pv}  (see \cite{Wiseman:2011by,Dias:2015nua} for a review), which first involves the selection of a reference metric $\bar{g}$ which has the same causal (regularity, horizon, and asymptotic) structure and, usually, the same symmetries as the solution we seek.

Our reference metric is based on the one used in \cite{Dias:2023rde}, which is a carefully chosen combination of the $\Lambda=0$ Bach-Weyl solution with two identical black holes \cite{BachWeyl(Republication):1922,Israel1964,Emparan:2001bb} and the static patch of de Sitter space.  Like the reference metric in \cite{Dias:2023rde}, ours has a $\mathbb Z_2$ symmetry, temporal and azimuthal Killing vectors $k=\partial/\partial t$ and $m=\partial/\partial \phi$, two black hole horizons, and a cosmological horizon.  Our main modification is the addition of a cross term $\dd t \dd\phi$ that introduces the angular velocities $\Omega _{+}^{(i)}$ ($i=1,2$) to the black holes in a frame where the de Sitter horizon is not rotating.  The $\mathbb Z_2$ symmetry forces the spin axes of the binaries to be aligned or anti-aligned.  Further details on the reference metric can be found in the Supplementary Material.

With a reference metric, we then solve the Einstein-DeTurck equation
\begin{equation} \label{eq:deturck}
R_{ab}-\nabla_{(a}\xi_{b)}=\frac{3}{\ell^2}g_{ab}\,,
\end{equation}
where $g_{ab}$ and $R_{ab}$ are the metric and its Ricci tensor, $\xi^a \equiv g^{bc}\left[\Gamma^a_{bc}(g)-\Gamma^a_{bc}(\bar{g})\right]$, and $\Gamma$ is the Christoffel connection.  Unlike the Einstein equation (without proper gauge fixing), the equation \eqref{eq:deturck} is elliptic \cite{Headrick:2009pv,Wiseman:2011by,Figueras:2011va,Dias:2015nua}, and yields a well-posed boundary problem with suitable boundary conditions.  In our case, boundary conditions are set by regularity and symmetry requirements.  The aligned $\Omega _{+}^{(2)}=\Omega _{+}^{(1)}$ and anti-aligned $\Omega _{+}^{(2)}=-\Omega _{+}^{(1)}$ cases have the same boundary conditions except at the $\mathbb{Z}_2$ symmetry plane, where the $g_{t\phi}$ metric component has Neumann and Dirichlet boundary conditions, respectively.

A solution to \eqref{eq:deturck} solves the Einstein equation only when $\xi=0$ (the solution will then be in the gauge $\xi=0$).  In certain cases, solutions to the Einstein-DeTurck equation are guaranteed to satisfy $\xi=0$ \cite{Figueras:2011va,Figueras:2016nmo}.  However, there are no such guarantees when $\Lambda>0$.  Fortunately, ellipticity guarantees local uniqueness. That is, solutions with $\xi=0$ cannot be arbitrarily close to those with $\xi\neq0$, and thus we can monitor the norm $\xi^a\xi_a$ to verify that our numerical discretization converges in the continuum to a true Einstein solution (see Supplementary Material).

\subparagraph{Results.}
With our symmetry requirements, the rotating de Sitter binaries form a two-parameter family.  Numerically, we take these parameters to be the horizon temperature $T_+\ell$ and angular velocity $\Omega _{+} \ell$.  Because scanning the full parameter space is computationally expensive, we leave a thorough sweep to future work. Instead we focus our attention on solutions with a fixed temperature $T_+\ell$, and varying $\Omega _{+} \ell$.  Note that the Newton-Hooke regime occurs for $T_+\ell \gg 1$ (\emph{i.e.\!} very small black holes).

\begin{figure}
    \centering
    \includegraphics[width=0.36\textwidth]{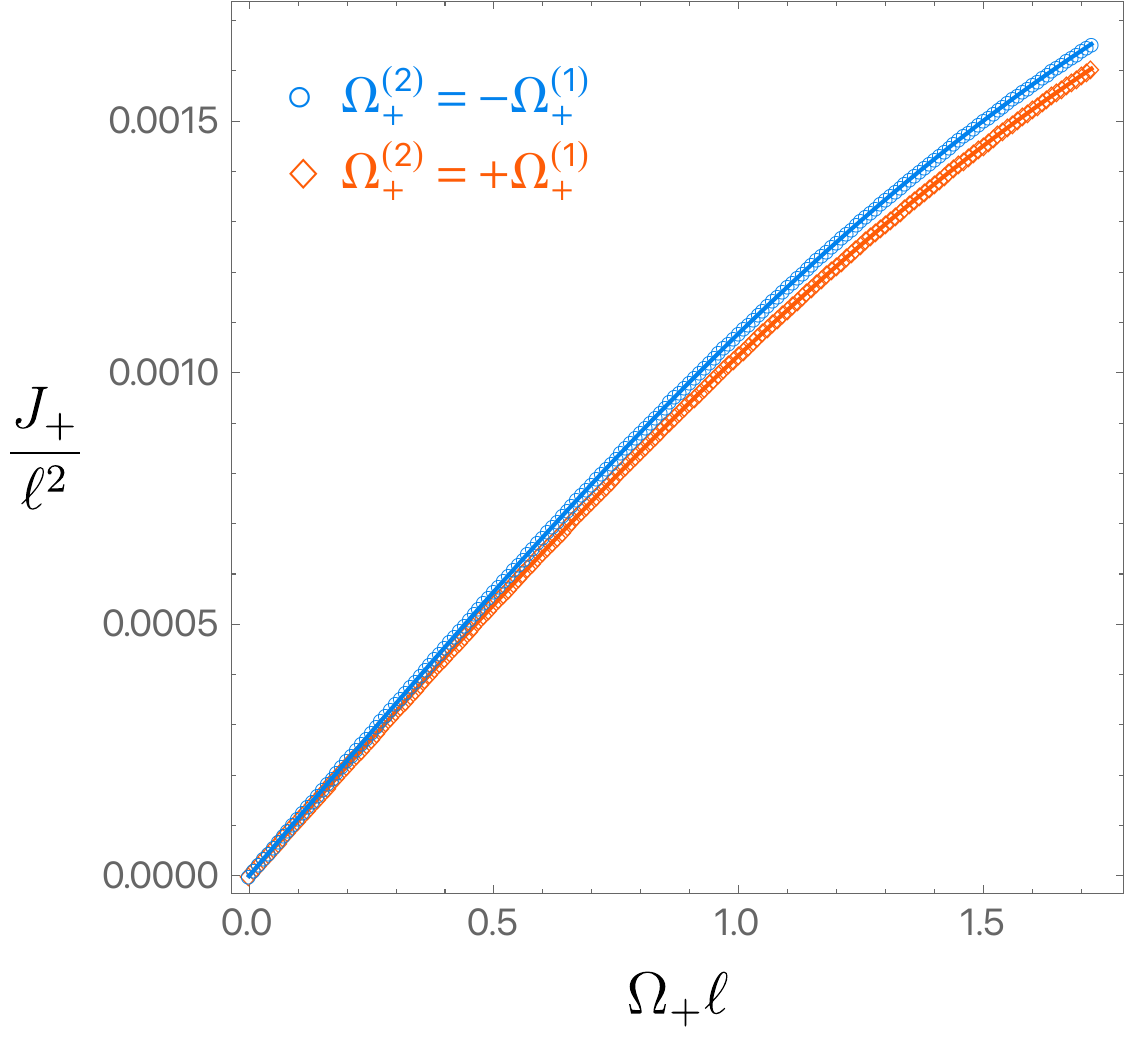}
    \caption{Black hole angular momentum of one of  binary's black hole versus its angular velocity for binaries with $T_+\ell=\frac{3.75}{2\pi}$.}
    \label{fig:JOmega}
\end{figure}

In Fig.~\ref{fig:JOmega}, we show the Komar angular momentum $J_+/\ell^2$ of one of the black holes versus its angular velocity $\Omega _{+} \ell$.  The highest angularity velocity we have reached is $\Omega _{+} \ell=1.72$.  We see no evidence of pathologies or singularities (see Supplementary Material for curvature invariants), so it is possible that these binaries exist for higher $\Omega _{+}$ (however, it cannot be ruled out that solutions stop existing  at a critical $\Omega_+\ell$ due to a lack of bound states).

We show both aligned and anti-aligned results, but they do not differ much.  We see that the angular momentum increases with angular velocity.  If solutions exist for higher $\Omega _{+}$, we would expect that the angular momentum will reach a maximum, as the same occurs for constant-temperature Kerr de Sitter black holes. A polynomial extrapolation gives a peak around $\Omega _{+} \ell=3$ \footnote{In a forthcoming publication \cite{DSW-chargedBinary}, we construct charged de Sitter binaries, and we see a maximum charge as the electric potential increases.}.

\begin{figure}[ht]
    \centering
    \includegraphics[width=0.36\textwidth]{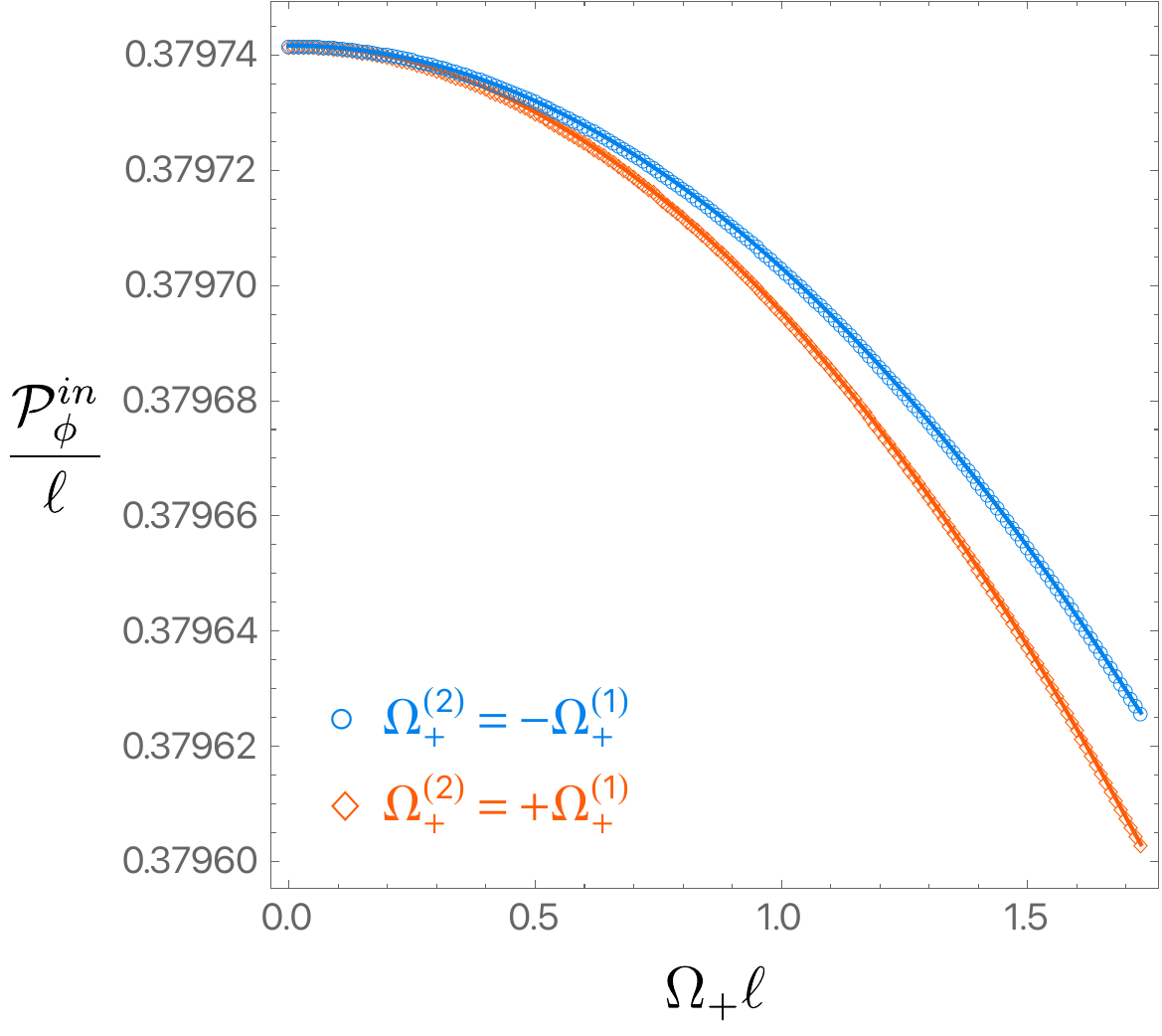}
    \caption{Proper distance along the symmetry axis between the two black holes, $\mathcal{P}_{\phi}^{in}/\ell$, versus angular velocity $\Omega_+\ell$ for $T_+\ell =\frac{9}{\pi}$.}
    \label{fig:proper_distanceInT9}
\end{figure}

We now discuss the proper distance $\mathcal{P}_{\phi}^{in}$ along the $\phi$ symmetry axis between the black hole horizons as a function of $\Omega _{+} \ell$. 
This is shown in  Fig.~\ref{fig:proper_distanceInT9} for a temperature $T_+\ell =\frac{9}{\pi}$ that seems sufficiently large to expect qualitative matching with Newton-Hooke expectations.  We see that, in agreement with the analysis of the Newton-Hooke relation \eqref{eq:NHpred}, by increasing spin the black holes move closer together with the aligned binary always having a smaller distance.

We now discuss the thermodynamics of the system.  In our case, the first law of black hole mechanics reads \cite{Gibbons:1977mu}
\begin{equation}
    2 T_+ \,\mathrm{d}S_+ +2\Omega_+  \,\mathrm{d}J_+ = -T_c\,\mathrm{d}S_c\,
\end{equation}
where $T_c$ is the temperature of the cosmological horizon, and $S_c$ is its entropy. 
We checked that our data obeys this law to within $0.01\%$.

\begin{figure}
    \centering
    \includegraphics[width=0.45\textwidth]{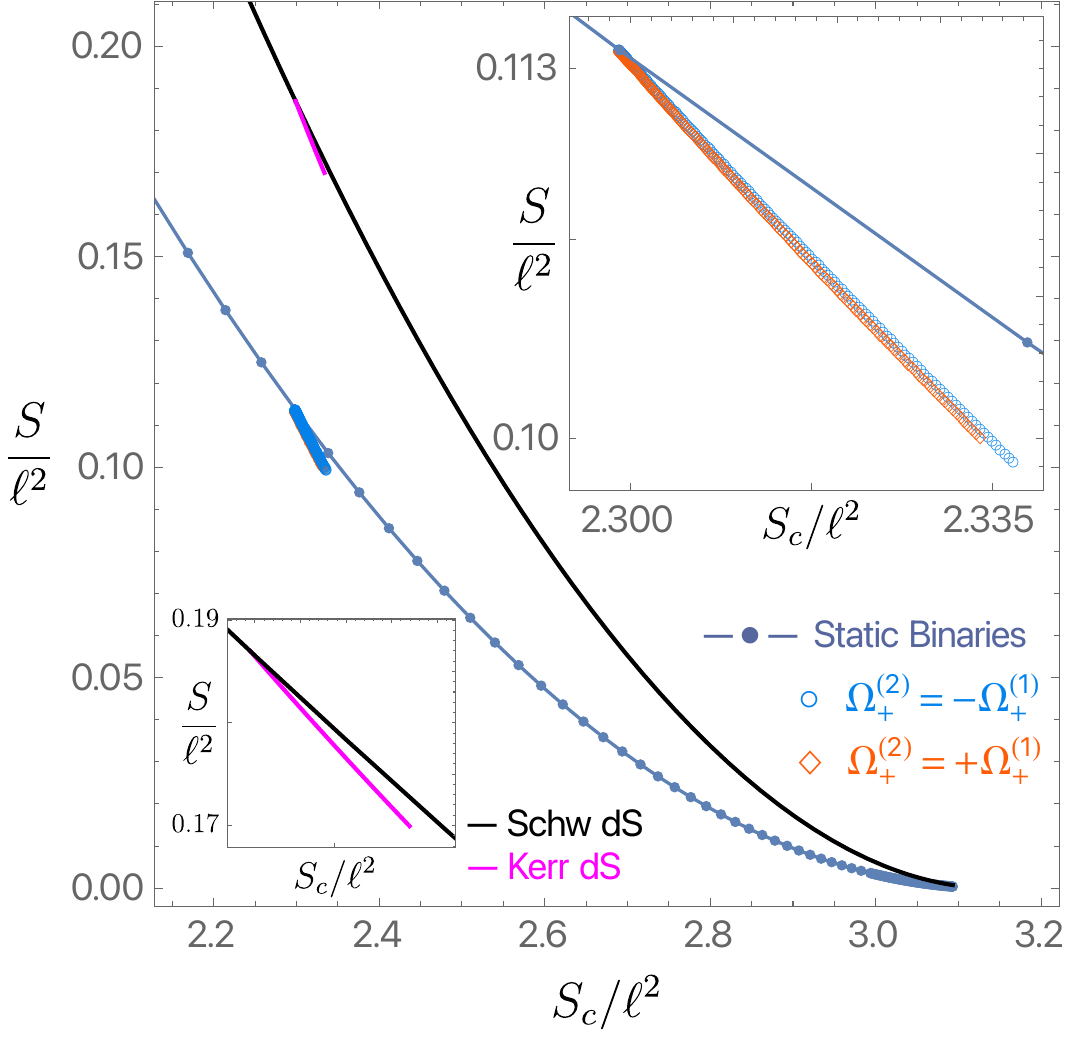}
    \caption{Projection of known asymptotically de Sitter (dS) static and stationary solutions in the plane that displays the total black hole entropy $S/\ell^2$ versus the cosmological horizon entropy $S_c/\ell^2$. The inset plots are zoom-ins of regions of particular interest as discussed in the text. The spinning binaries have $T_+\ell=\frac{3.75}{2\pi}$.}
    \label{fig:micro}
\end{figure}

\begin{figure*}[th]
    \centering
    \includegraphics[width=0.5\textwidth]{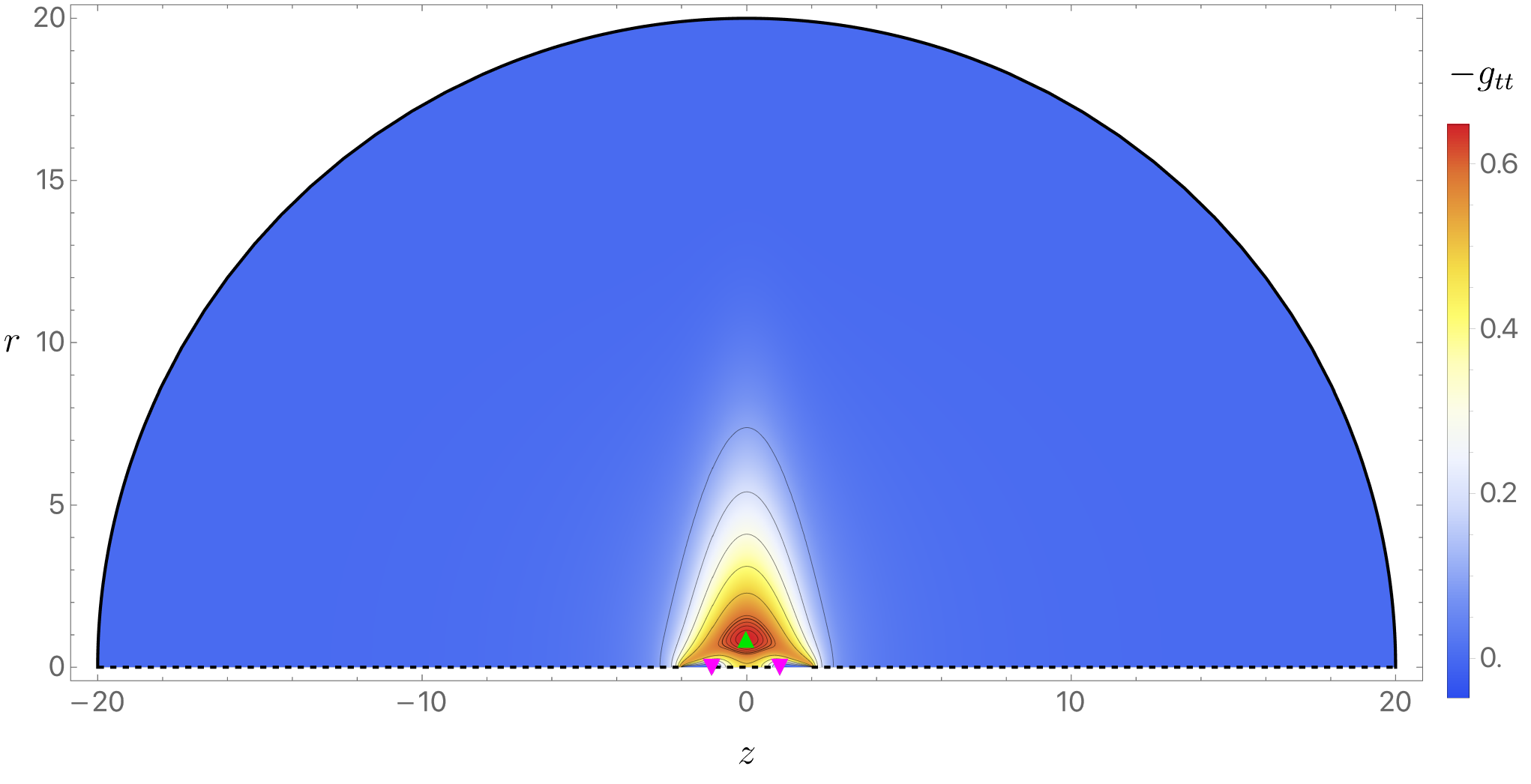}\hspace{0.5cm}
    \includegraphics[width=0.4\textwidth]{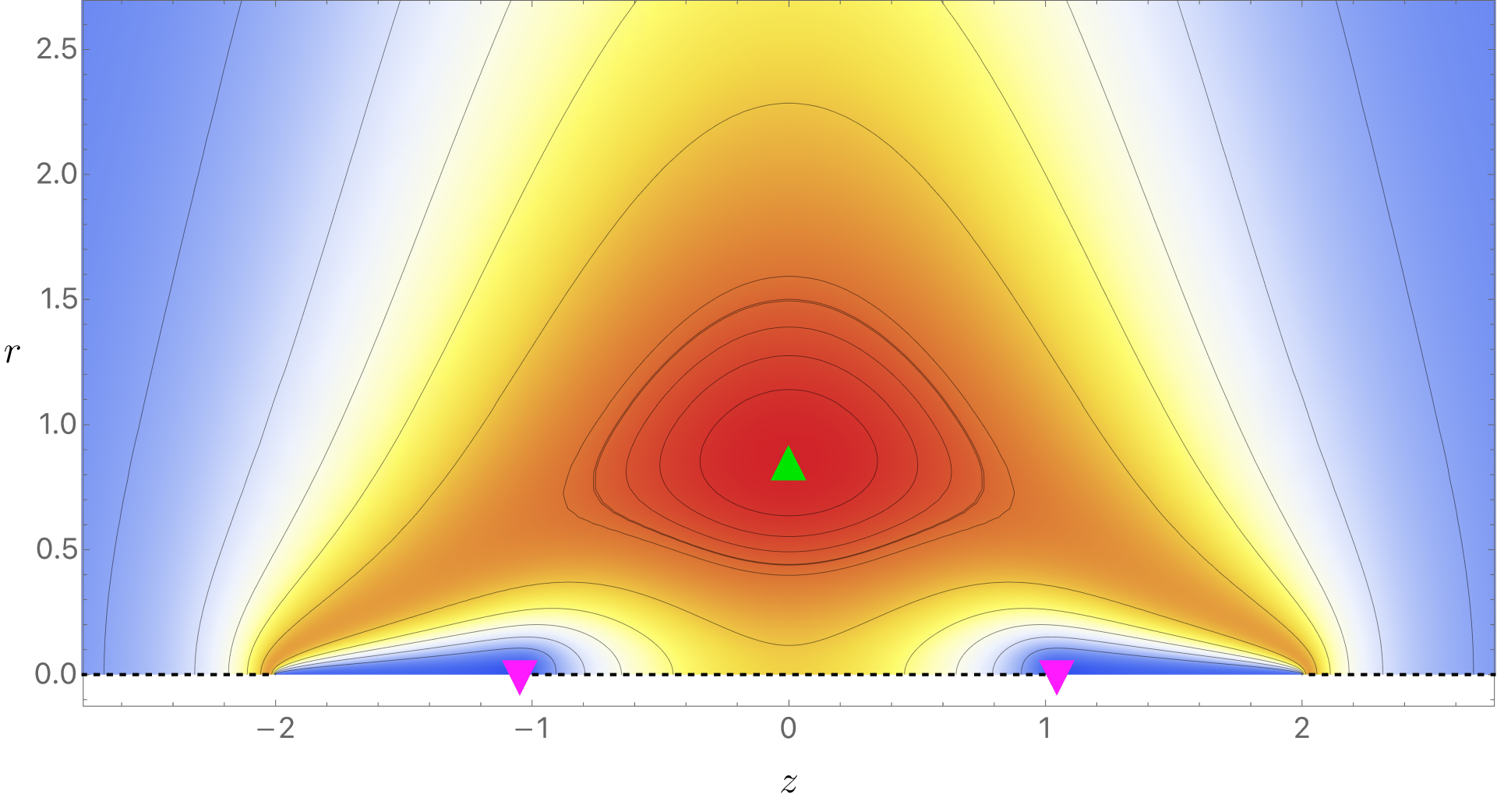}
    \caption{Contour plot showing the level sets of the gauge invariant metric component $-g_{tt}$.  The cosmological horizon is the outer solid black semicircle.  The two black hole horizons are located along $r=0$ at $z \in [-1,-2]$ and $z \in [1,2]$, and the outer and inner axes are the dashed black lines at $r=0$. The inset plot is a zoom around the two black holes. The green triangle is where $-g_{tt}$ takes its maximum value and the magenta inverted-triangle is where $-g_{tt}$ takes its minimum (negative) value. }
    \label{fig:level}
\end{figure*}

In Fig.~\ref{fig:micro}, we show the total black hole entropy $S/\ell^2$ against the cosmological horizon entropy $S_c/\ell^2$ \cite{Gibbons:1977mu}.  Again, we are only showing the spinning de Sitter binaries for the constant temperature $T\ell=\frac{3.75}{2\pi}$.  For reference, we also included the corresponding values for Schwarzschild de Sitter, Kerr de Sitter, and the static binaries from \cite{Dias:2023rde}. In the Kerr case, we only show black holes with the same total angular momentum as the aligned de Sitter binaries. The anti-aligned binaries have zero total angular momentum, so should be compared with Schwarzschild and the static binaries.  We see that in all cases, the spinning binaries have less entropy $S$ than the corresponding Schwarzschild or Kerr black holes, and so do not dominate the microcanonical ensemble.

From a dynamical perspective, the cosmological horizon entropy does not necessarily stay constant during evolution.  But, both horizons satisfy a second law, so it cannot decrease during evolution.  We note that our maximal entropy in Fig.~\ref{fig:micro} is monotonically decreasing with $S_c$.  That is, the maximum entropy $S_{\mathrm{max}}(S_c)$ is larger than any entropy $S_{\mathrm{max}}(S_c')$ for $S_c'>S_c$. This suggests that the endpoint that maximizes the entropy is indeed the Schwarzschild de Sitter black hole and also suggests its stability.  Similar arguments can be made in asymptotically flat space, where energy and angular momentum  can be radiated to infinity, and hence are not conserved during evolution.

\subparagraph{Conclusions.} We constructed the first examples of stationary, spinning black binary solutions of general relativity with a positive cosmological constant where the gravitational, cosmological and spin-spin interactions combine to produce an equilibrium configuration. We have found these solutions well away from the regime where Newton-Hooke theory is valid, but the properties of our binaries match the Newton-Hooke expectations.

We point out that, like in \cite{Dias:2023rde}, the available uniqueness theorems \cite{LEFLOCH20101129,Borghini:2019msu,Katona:2023vtq,Katona:2024oah} do not actually forbid the existence of these de Sitter binaries.  This is largely because such theorems make assumptions about the level sets of the gauge-invariant metric component $-g_{tt}$.  Fig.~\ref{fig:level} shows such level sets for a binary with $\Omega_+^{(2)}\ell=\Omega_+^{(1)}\ell=1.7$ (other solutions are qualitatively similar).  The location of the maximum of $-g_{tt}$ is marked by a green triangle.  The fact that this maximum is a point in this plot and not a line demonstrates that the assumptions in these theorems are violated (we refer the reader to \cite{Dias:2023rde} for a more detailed discussion on this matter).

For completeness, we also show the minima of $-g_{tt}$ marked by the two magenta inverted-triangles.  These minima take negative values because of the ergoregion around each black hole (where energy and angular momentum can be extracted from the Penrose process or superradiant scattering).

It is clear that black hole uniqueness is strongly violated in de Sitter, unlike in flat space.  Indeed, we have an infinite non-uniqueness. For example, in the anti-aligned case, we have a whole continuous two-parameter family of solutions with zero total angular momentum.  Therefore, there is no analogous no-hair conjecture \cite{Ruffini:1971bza} for pure gravity with a $\Lambda>0$.

Although the solutions we have found do not dominate the microcanonical ensemble, this does not necessarily mean that they are dynamically unstable.  Because spin-spin interactions act on shorter length scales than that of the Newtonian potential, they may provide a mechanism for stabilizing the binary in some regions of parameter space, much like the stability of molecules \cite{Bohr1913a,Bohr1913b,Bohr1913c,Svidzinsky2014}. It would be interesting to see if this intuition also holds within general relativity, which would require a proper stability analysis of our numerical spinning binaries (or their extensions in the parameter space).

We have left a more thorough search of parameter space to future work.  This could include higher angular velocities as well as a sweep over the full 2-parameter family.  It would also be interesting to drop the $\mathbb{Z}_2$ symmetry restrictions and consider unequal black holes with unequal spins.  In the case where one of the black holes is much smaller than the other, the Mathisson-Papapetrou-Dixon's formalism \cite{mathisson2010republication,papapetrou1951spinning,corinaldesi1951spinning,Dixon:1970zza,dixon1973definition,Dixon:1974xoz,Wald:1972sz} can be used.   Work in this direction is underway.

Finally, we note that our solutions have spin but no orbital angular momentum. Adding orbital angular momentum will also create quadrupole moments that emit gravitational radiation and hence, the system will not be stationary.  Nevertheless, our solutions can provide initial data that satisfy the elliptic constraint equations of the time evolution problem, that ultimately leads to the black hole binary merger and associated gravitational (and electromagnetic) wave emission.

\begin{acknowledgments}
\subparagraph{Acknowledgments.}
 O.D. acknowledges financial support from the STFC ``Particle Physics Grants Panel (PPGP) 2018" Grant No.~ST/T000775/1 and PPGP 2020 grant No.~ST/X000583/1. O.D.'s research was also supported in part by the YITP-ExU long-term workshop {\it Quantum Information, Quantum Matter and Quantum Gravity (QIMG2023)}, Yukawa Institute for Theoretical Physics, Kyoto Univ., where part of this work was completed. J.~E.~S. has been partially supported by STFC consolidated grant ST/T000694/1.  The authors acknowledge the use of the IRIDIS High Performance Computing Facility, and associated support services at the University of Southampton, for the completion of this work.
\end{acknowledgments}


\widetext
\appendix
\section*{Supplementary Material}

\section{Metric Ansatz for Spinning Black Binaries in de Sitter}

In the Einstein-DeTurck formalism, \eqref{eq:deturck} requires a reference metric $\bar{g}$ which has the same casual (asymptotic and horizon) structure and, whenever possible, also the same symmetries as the solution we seek.  We will follow the same procedure in \cite{Dias:2023rde} to obtain the reference metric, making a few  modifications to accommodate spin.

The reference metric in \cite{Dias:2023rde} was built from a careful combination of the Bach-Weyl \cite{BachWeyl(Republication):1922} (i.e., the Israel-Khan solution \cite{Israel1964} with two black holes) and de Sitter space.  The Bach-Weyl solution is a closed form, asymptotically flat solution of the Einstein equation that consists of two non-spinning black holes held together by a conical singularity.  The overall strategy is to combine this singular binary with the de Sitter horizon in hopes of finding a de Sitter solution with a conical singularity, and then remove the conical singularity.

The Bach-Weyl solution is usually written in Weyl coordinates $(r,z)$ because the resulting Einstein equation in this gauge can be expressed in an integrable form.  These coordinates have a ``rod-structure" so that the coordinate line $r=0$ includes both inner and outer segments of the $\phi$ symmetry axis as well as the black hole horizons.  This rod-structure is inconvenient for numerical purposes, largely because it introduces more coordinate singularities and makes the application of boundary conditions difficult. Instead, in \cite{Dias:2023rde}, a transformation
 \begin{equation}\label{xydef}
z=\frac{x\sqrt{2-x^2}\sqrt{(1-y^2)^2+k^2y^2(2-y^2)}}{(1-y^2)^2+k^2x^2(2-x^2)y^2(2-y^2)}\;,\qquad r=\frac{(1-x^2)\sqrt{1-k^2x^2(2-x^2)}y\sqrt{2-y^2}(1-y^2)}{(1-y^2)^2+k^2x^2(2-x^2)y^2(2-y^2)}\;,
\end{equation}
was used to map both horizons, as well as the outer and inner segments of the axis into the four sides of a coordinate rectangle $(x,y)$. 
The Bach-Weyl solution \cite{BachWeyl(Republication):1922,Israel1964} then takes the form \cite{Dias:2023rde}
\begin{equation}
\label{IKxy}
\dd s^2=\ell^2\left\{-f\dd t^2+\frac{\lambda^2}{m^2 \Delta_{xy}^2}\left[p^2\left(\frac{4\dd x^2}{(2-x^2)\Delta_x}+\frac{4\dd y^2}{(2-y^2)\Delta_y}\right)+y^2(2-y^2)(1-y^2)^2\dd \phi^2\right]\right\}\;,
\end{equation}
where $\ell$ is an arbitrary dimensionful length scale that we have introduced for later use in de Sitter, and
\begin{align}\label{IKxy:Aux}
& \Delta_x=1-k^2x^2(2-x^2)\;,\qquad \Delta_y=1-(1-k^2)y^2(2-y^2)\;,\qquad \Delta_{xy}=(1-y^2)^2+k^2x^2(2-x^2)y^2(2-y^2)\;, \\
& f=(1-x^2)^2\Delta_x m^2\;,\quad
 p=\frac{k\left(1+\sqrt{\Delta_y}\right)^2}{(1+k)^2}\;,\quad m=\frac{k\Big[1-(1-k)y^2(2-y^2)+\sqrt{\Delta_y}\Big]}{(1-k)\Delta_x(1-y^2)^2+(k+\sqrt{\Delta_y})\Big[\Delta_x+(1-k)(\sqrt{\Delta_{xy}}-1)\Big]}\;.   \nonumber
\end{align}
All functions $f$, $\Delta_x$, $\Delta_y$, $p$, and $m$ are smooth and positive definite in the domain.  $\Delta_{xy}$ vanishes at $(x,y)=(0,1)$ (asymptotic infinity), and is positive and smooth otherwise.  The solution is parametrized by $k\in (0,1)$.

To eventually accommodate a cosmological horizon, we also consider the polar Weyl coordinates $(\rho,\xi)$ defined by:
\begin{equation}\label{rhoxidef}
z=\rho\,\xi\sqrt{2-\xi^2}\;,\qquad r=\rho(1-\xi^2)\;,
\end{equation}
where the Bach-Weyl solution \cite{BachWeyl(Republication):1922,Israel1964}
takes the form \cite{Dias:2023rde}
\begin{equation}\label{IKpolar}
\dd s^2=\ell^2\left\{-f\dd t^2+\frac{\lambda^2h}{f}\left[\dd \rho^2+\rho^2\left(\frac{4\dd\xi^2}{2-\xi^2}+\frac{(1-\xi^2)^2}{h}\dd\phi^2\right)\right]\right\}\;,
\end{equation}
with
\begin{align}\label{IKpolar:Aux}
& R_\pm=\sqrt{\rho^2+\frac{1}{k^2}\pm\frac{2}{k}\rho\,\xi\sqrt{2-\xi^2}}\;,\qquad r_\pm=\sqrt{\rho^2+1\pm 2 \rho\,\xi\sqrt{2-\xi^2}}\;, \nonumber\\
& f=\left(\frac{k(R_++r_+)-(1-k)}{k(R_++r_+)+(1-k)}\right)\left(\frac{k(R_-+r_-)-(1-k)}{k(R_-+r_-)+(1-k)}\right)\;,\nonumber\\
& h=\left(\frac{\rho^2+\tfrac{1}{k}[1+(1+k)\rho\,\xi\sqrt{2-\xi^2}]+R_+r_+}{2R_+r_+}\right)\left(\frac{\rho^2+\tfrac{1}{k}[1-(1+k)\rho\,\xi\sqrt{2-\xi^2}]+R_-r_-}{2R_-r_-}\right)   \\
&\qquad \times\left(\frac{\rho^2-1+r_+r_-}{\rho^2-\tfrac{1}{k}[1+(1-k)\rho\,\xi\sqrt{2-\xi^2}]+r_+R_-}\right)\left(\frac{\rho^2-(1/k^2)+R_+R_-}{\rho^2-\tfrac{1}{k}[1-(1-k \,\rho\,\xi\sqrt{2-\xi^2})]+R_+r_-}\right)\;. \nonumber
\end{align}
Note that $h$ and $f$ approach unity when $\rho\to\infty$, where the spacetime becomes asymptotically flat.

Now we separately consider de Sitter space, which can be written in isotropic coordinates \cite{Dias:2023rde}
\begin{equation}\label{dSisotropic}
\dd s^2=\frac{\ell^2}{g_+^2}\left\{-g_-^2\dd t^2+\lambda^2\left[\dd \rho^2+\rho^2\left(\frac{4\dd\xi^2}{2-\xi^2}+(1-\xi^2)^2\dd\phi^2\right)\right]\right\}\;, \qquad \hbox{with} \:\: g_\pm=1\pm\frac{\lambda^2\rho^2}{4},
\end{equation}
where  $\ell$ is the de Sitter radius $\ell=\sqrt{\Lambda/3}$. In these coordinates, the de Sitter horizon has a constant temperature of $T_c=1/(2\pi)$.  In \eqref{dSisotropic}, as in \eqref{IKpolar}, $\lambda$ is a gauge parameter that simply scales the radial coordinate $\rho$.  

The form \eqref{dSisotropic} for de Sitter space is suggestively similar to the Bach-Weyl solution \eqref{IKpolar} in polar Weyl coordinates.  Aside from some factors of $f$ and $h$ (which approach unity at large $\rho$), the only differences are that de Sitter in isotropic coordinates has an overall conformal factor of $1/g_+^2$ and a factor of $g_-^2$ in the $\dd t^2$ term whose zero defines the de Sitter horizon. This effectively constrains the $\rho$ coordinate in our integration domain to $\rho<2/\lambda$. We will make use of these similarities below.

Now we can form a reference metric as was done in \cite{Dias:2023rde}.
To do so note that from \eqref{xydef} and \eqref{rhoxidef}, we can find an explicit coordinate transformation between the $(x,y)$ coordinates and the polar Weyl $(\rho,\xi)$ coordinates:
\begin{subequations}\label{trans}
\begin{align}
\rho&=\frac{\sqrt{y^2(2-y^2)+x^2(2-x^2)(1-y^2)^2}}{\sqrt{(1-y^2)^2+k^2x^2(2-x^2)y^2(2-y^2)}}\;,\\
\xi&=\sqrt{1-\frac{(1-x^2)y\sqrt{2-y^2}(1-y^2)\sqrt{1-k^2x^2(2-x^2)}}{\sqrt{y^2(2-y^2)+x^2(2-x^2)(1-y^2)^2}\sqrt{(1-y^2)^2+k^2x^2(2-x^2)y^2(2-y^2)}}}\;.
\end{align}
\end{subequations}
which is needed in what follows.

With these ingredients in place, we can finally describe our reference metric:
\begin{align}\label{Spin:refG}
\dd s^2_{\mathrm{ref}}&=\frac{\ell^2}{g_+^2}\left\{-fg_-^2\,F\,\dd t^2+\frac{\lambda^2}{m^2 \Delta_{xy}^2}\left[p^2\left(\frac{4\dd x^2}{(2-x^2)\Delta_x}+\frac{4\dd y^2}{(2-y^2)\Delta_y}\right)+y^2(2-y^2)(1-y^2)^2 \,s\, \left(\dd\phi- \beta \,g_-^2 w \,\dd t \right)^2\right]\right\}\nonumber\\
&=\frac{\ell^2}{g_+^2}\left\{-fg_-^2\,F\,\dd t^2+\frac{\lambda^2h}{f}\left[\dd \rho^2+\rho^2\left(\frac{4\dd\xi^2}{2-\xi^2}+\frac{(1-\xi^2)^2}{h}\,s\,  \big(\dd\phi - \beta \,g_-^2 w \,\dd t \big)^2\right)\right]\right\}\;,
\end{align}
with several functions that appear in theses metrics defined in \eqref{IKxy:Aux} or \eqref{IKpolar:Aux}.  We will describe the remaining functions further below. The equality between the first and second lines of~\eqref{Spin:refG} (here and in the remainder of this section) is understood to be through the coordinate transformation \eqref{trans}.  We will use $(x,y)$ coordinates in the region near the black holes and inner segment of the axis, and the $(\rho,\xi)$ coordinates near the cosmological horizon. The DeTurck metric \eqref{Spin:refG} is a simple upgrade of the static DeTurck reference metric (B5) of \cite{Dias:2023rde} where we have simply introduced a $g_{t\phi}$ term to accommodate the spins of the black holes. More concretely, we have simply taken (B5) of \cite{Dias:2023rde}  and made the replacement $\dd \phi^2 \to \left(\dd\phi- \beta g_-^2 w \,\dd t \right)^2$ (which also introduces a new function $w$ defined below). This reference metric depends on four parameters, $\lambda,k,\alpha$ and $\beta$ whose physical interpretation is given below.

We only made four changes to the Bach-Weyl  (Israel-Khan) solution to arrive at the reference metric~\eqref{Spin:refG}. The first is the inclusion of a conformal factor $1/g_+^2$ to enable the matching to de Sitter spacetime \eqref{dSisotropic} in isotropic coordinates in the far-region.  The second is the inclusion of a function $s$ in the $\dd\phi^2$ term with a conical parameter $\alpha$ that we will adjust to remove any conical singularities. This is achieved with the choices
\begin{equation}\label{Spin:alphaspecial}
s=1-\alpha(1-y^2)^2\;, \qquad \alpha =\frac{(1-k)^2 \left(k^2+6 k+1\right)}{(k+1)^4}\,.
\end{equation}
The third change is the inclusion of a factor of $g_-^2\,F$ in the $\dd t^2$ term which introduces a cosmological horizon at $\rho=2/\lambda$ ($F$ is a complicated function that  will be given further below in \eqref{Spin:Fdef}).
The fourth change is the replacement $\dd \phi^2 \to \left(\dd\phi- \beta g_-^2 w \,\dd t \right)^2$ where $w$ must satisfy  $g_-^2 w|_{x=\pm 1}=1$  but can otherwise be freely chosen. We choose
\begin{equation}
w=\frac{1}{ \left(1-x^2\right)^2+g_-^2 }\,.
\end{equation}
This fourth change introduces an angular velocity at the horizons $x=\pm 1$ (which is given by the novel parameter $\beta$) but leaves the cosmological horizon (where  $g_-$ vanishes) with zero rotation.

We have some freedom to choose the function $F$, but the choice is delicate \cite{Dias:2023rde}.  For numerical reasons, we want $F$ to be smooth both in $(x,y)$ and $(\rho,\xi)$ coordinates.  The DeTurck method further requires that $F$ preserves the regularity of both the black hole and cosmological horizons \cite{Headrick:2009pv,Wiseman:2011by,Dias:2015nua}. In addition, to aid in finding a solution using a Newton-Raphson algorithm, it would be convenient to also choose $F$ to match physical expectations in certain limits.  Specifically, we expect that when the cosmological horizon is large compared to other length scales (\emph{i.e.\!} $\lambda\ll1$), the spacetime near the cosmological horizon should approach de Sitter and the spacetime closer to the origin should be approximately described by the Bach-Weyl solution (when $\alpha=0$ and $\beta=0$). As detailed in \cite{Dias:2023rde}, these requirements are achieved if we choose $F$ to be \cite{Dias:2023rde}
\begin{equation}\label{Spin:Fdef}
F=\frac{G}{f+g_-^2G-fg_-^2G}\;,\qquad \hbox{with }\quad G=\frac{\tfrac{\hat h}{\hat f}(1-x^2)y^2(2-y^2)+g_-^2}{(1-x^2)y^2(2-y^2)+g_-^4}\;,
\end{equation}
where $\hat f$ and $\hat h$ are any smooth, positive definite functions that agree with $f$ and $h$, respectively at $\rho=2/\lambda$.  To choose $\hat f$ and $\hat h$, we first take the expressions for $f$ and $h$ as written in \eqref{IKpolar:Aux}, and treat them as functions $f(\rho,z)$ and $h(\rho,z)$. We then set $\hat f(\rho,\xi)=f(2/\lambda,\rho\xi\sqrt{2-\xi^2})$ and similarly for $\hat h$.  Note that we cannot use a choice like $\hat f(\rho,\xi)=f(\tfrac{2}{\lambda},\tfrac{2}{\lambda}\xi\sqrt{2-\xi^2})$ as it is not smooth in the $(x,y)$ coordinates at $x=0,y=1$.

The reference metric  \eqref{Spin:refG} is parametrized by $\lambda$, $k$, $\alpha$ and $\beta$.
This metric describes a solution that is everywhere regular when the conical parameter $\alpha$ to be given by \eqref{Spin:alphaspecial}, with black hole horizons at $x=\pm 1$. These  are Killing horizons generated by the Killing vector field $K= \partial_t+\Omega _{+}  \partial_\phi$, \emph{i.e.\!}  $|K|_{x=\pm 1}=0$ (where we use the fact that $f=(1-x^2)^2\Delta_x m^2$ and  $g_-^2 w|_{x=\pm 1}=1$).  The temperature $T_+$ and angular velocity $\Omega _{+} $ are given by
\begin{equation}\label{Spin:IKtemp}
    T_+ = \frac{1}{2\pi}\frac{k(1+k)}{4\lambda(1-k)}\,, \qquad   \Omega _{+} = \beta\,,
\end{equation}
where $\lambda \in (0,\infty)$ and $k\in (0,1)$. Note that because of the overall factor of $\ell$ in the metric, $T_+$ and $\Omega_+$ as expressed here are dimensionless. Any $k$ and $\lambda$ that give the same $T_+$ are physically equivalent. Effectively, $\lambda$ is a gauge parameter that only sets an overall scale (it fixes the cosmological horizon location to be at $\rho=2/\lambda$) and $k$ then fixes the black hole temperature. Solutions described by \eqref{Spin:refG} also have a cosmological horizon at $\rho=2/\lambda$ (where $g_-$  vanishes) with temperature $T_c=1/(2\pi)$ and angular velocity $\Omega_c=0$. This is generated by the Killing vector field $\zeta= \partial_t+\Omega_c\partial_\phi=\partial_t$, \emph{i.e.\!}  $|\zeta|_{\rho=2/\lambda}=0$. Note that, unlike the event horizon, the cosmological horizon has vanishing angular velocity. The reference metric therefore only has two physical parameters, $T_+$ and $\Omega_+$ (or $k$ and $\beta$).

With a reference metric, we can finally write our metric ansatz for the spinning binary in de Sitter, which is essentially the most general ansatz compatible with the symmetries of the problem:
\begin{align}\label{Spin:G}
\dd s^2 &=\frac{\ell^2}{g_+^2}\Bigg\{-fg_-^2\,F\,\mathcal{T}\,\dd t^2+\frac{\lambda^2}{m^2 \Delta_{xy}^2}\Bigg[p^2\left(\frac{4 \mathcal{A}\,\dd x^2}{(2-x^2)\Delta_x}+\frac{4 \mathcal{B}}{(2-y^2)\Delta_y} \left(\dd y-x\,(1-x^2)\,y\,(2-y^2)(1-y^2)\mathcal{F}\,\dd x \right)^2\right) \nonumber\\
 & \hspace{5.3cm} +y^2(2-y^2)(1-y^2)^2 \,s\, \mathcal{S} \left(\dd\phi- \,g_-^2 w \,\mathcal{W} \,\dd t \right)^2\Bigg]\Bigg\}\nonumber\\
&=\frac{\ell^2}{g_+^2}\Bigg\{-fg_-^2\,F\,\widetilde{\mathcal{T}}\,\dd t^2+\frac{\lambda^2h}{f}\Bigg[\widetilde{\mathcal{A}}\,\dd \rho^2+\rho^2\Bigg(\frac{4\widetilde{\mathcal{B}}}{2-\xi^2}\left(\dd\xi -\xi \,(2-\xi^2)(1-\xi^2)\,\rho\, \widetilde{\mathcal{F}}\,\dd\rho \right)^2
 \nonumber\\
 & \hspace{4.8cm}
+\frac{(1-\xi^2)^2}{h}\,s \, \widetilde{\mathcal{S}}  \big(\dd\phi -  \,g_-^2 w\, \widetilde{\mathcal{W}}  \,\dd t \big)^2\Bigg)\Bigg]\Bigg\}\;.
\end{align}
where $\{\mathcal{Q}\}\equiv \{ \mathcal{T}, \mathcal{A}, \mathcal{B}, \mathcal{F}, \mathcal{S}, \mathcal{W}\}$ are unknown functions of $\{x,y\}$ and $\{\widetilde{\mathcal{Q}} \}\equiv\{ \widetilde{\mathcal{T}}, \widetilde{\mathcal{A}}, \widetilde{\mathcal{B}}, \widetilde{\mathcal{F}}, \widetilde{\mathcal{S}}, \widetilde{\mathcal{W}}\}$  are unknown functions of $\{\rho,\xi\}$. The known functions, already present in \eqref{Spin:refG} and defined in \eqref{IKxy:Aux} or \eqref{IKpolar:Aux}, should  be treated as scalars, transforming between coordinate systems as \eqref{trans}. Note that when we set $\{ \mathcal{T}=1, \mathcal{A}=1, \mathcal{B}=1, \mathcal{F}=0, \mathcal{S}=1, \mathcal{W}=1, \mathcal{W}=\beta\}$  (and similarly for the associated set of tilde functions $\{\widetilde{\mathcal{Q}}\}$), we recover the DeTurck reference metric \eqref{Spin:refG}.  The choice of our treatment of the factor $\beta$ allows us to more easily recover the static results \cite{Dias:2023rde} in the limit $\beta\to 0$.  

Most of the boundary conditions are determined by regularity, and we refer readers to Section V of the review \cite{Dias:2015nua} for a detailed discussion of such boundary conditions.  Many of these are set by including factors of $x$, $1-x$, $y$ and $1-y$ (and similar factors of $\xi$, $1-\xi$, $\rho$ and $1-\rho$ into the metric ansatz.  The $\mathbb{Z}_2$ reflection plane at $x=0$ ($\xi=0$) deserves a few further comments.  Our reference metric is even about this surface, which agrees with the desired solutions with aligned spins, for which we demand Neumann boundary conditions.  For the anti-aligned case, we take the same boundary conditions except for $\mathcal{W}$ ($\widetilde{\mathcal{W}}$), for which we choose a Dirichlet condition.  The fact that the reference metric does not itself satisfy this condition does not lead to any contradictions. This is indeed not a problem since this is happening at an axis of symmetry (not at a boundary of the spacetime) and the reference metric still has well defined parity (moreover, one explicitly checks that the de Turck norm indeed vanishes along this axis even with the Neumann boundary conditions).

We supplement the boundary conditions by a set of patching conditions to join the two different coordinate systems together. We defer a discussion of patching to the next section.

For fixed cosmological constant, our solutions are physically parametrised by $T_+$ and $\Omega_+$, but our solutions are parametrised by four parameters: $\alpha$, $\beta$, $\lambda$, $k$.  $\alpha$ is set according to \eqref{Spin:alphaspecial} to remove conical singularities, $\beta$ is uniquely determined by a choice of $\Omega_+$, but any combination of our parameters $\lambda$ and $k$ that give the same black hole temperature $T_+$ are physically equivalent.  After some trial and error, we find good numerical results by fixing $\lambda=1/10$ and using $k$ and $\beta$ to parametrize the solutions.

Our (aligned or anti-aligned) spinning de Sitter binaries are thus  a 2-parameter family of solutions parametrized by $T_+/T_c$ and $\Omega_+$. Any combination of $k$ and $\lambda$ that give the same black hole temperature $T_+$ are physically equivalent.  To collect our numerical data, we typically have fixed $\lambda=1/10$ and used $k$ and $\beta$ to parametrize our solutions.  We have tried different values of $\lambda$, but after trial and error, this value typically generated the best numerical results.

\section{Patching and Numerical Methods}\label{sec:patching}
In this section, we explain how we partition the domain of integration using patching techniques (reviewed  \emph{e.g.} in \cite{Dias:2015nua}). The solution we seek contains five boundaries: the inner segment of the axis ($\partial_{\phi}^{\rm in}$), the black hole horizon ($\mathcal{H}^+$), the outer segment of the axis ($\partial_{\phi}^{\rm out}$), the cosmological horizon ($\mathcal{H}^c$), and the plane of $\mathbb{Z}_2$ symmetry.  Near the black hole event horizon we use $(x,y)$ coordinates, while near the cosmological horizon we use $(\rho,\xi)$ coordinates, and these coordinates at patched together.

\begin{figure}
    \centering
    \includegraphics[width=1\textwidth]{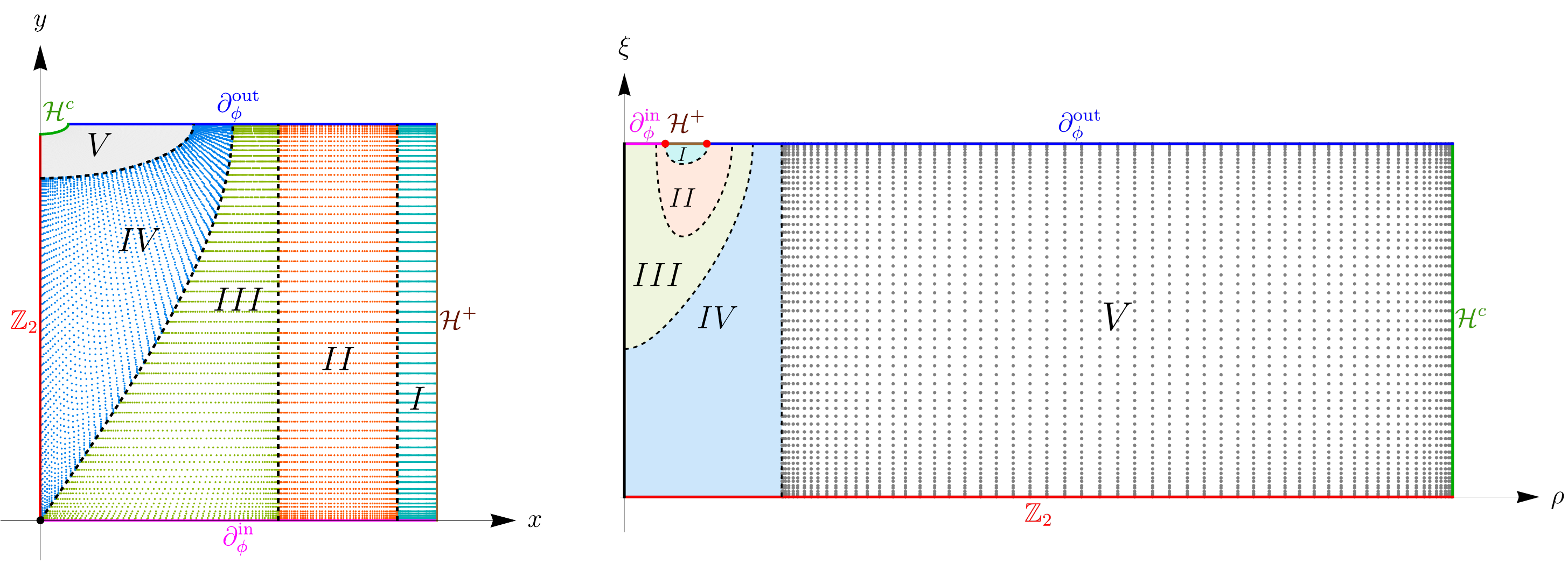}
    \caption{Illustration of the patches used in our numerical construction. This is for $\lambda=0.1$, $k=0.5$ and $\rho_0, x_0,\chi_0,\chi_1$ as in~\eqref{grid}.}
    \label{fig:map}
\end{figure}

For the results presented in this Letter, we used a total of five patches $-$ $I$, $II$,$III$, $IV$ and $V$ $-$ with each of these having four boundaries: see Fig.~\ref{fig:map}. Patches $I, II, III$ and $IV$ are defined in $(x,y)$ coordinates, and patch $V$ in $(\rho,\xi)$ coordinates. The patching boundary (dashed line in  Fig.~\ref{fig:map}) between patches $I$ and $II$ is given by constant $x=\chi_1$, and between patches $II$ and $III$ it is given by  constant $x=\chi_0$. The patching boundary between patches $III$ and $IV$ is given by $x=x_0 y\sqrt{2-y^2}$. Finally, the patching boundary between patch $IV$ and patch $V$ is given by $\rho=\rho_0$.
We fix the grid parameters $\rho_0, x_0, \chi_0$, and $\chi_1$ by:
\begin{equation}\label{grid}
\rho_0=\frac{1}{10} \left(\frac{2}{\lambda }-\frac{1}{k}\right)+\frac{1}{k}\,\quad x_0=\frac{7}{10} \left(1-\sqrt{1-\sqrt{1-\frac{1}{k^2 \rho _0^2}}}\right)+\sqrt{1-\sqrt{1-\frac{1}{k^2 \rho _0^2}}},\quad \chi_0=0.6 \quad\text{and}\quad \chi_1=0.9 \,,
\end{equation}

We apply the numerical methods detailed in \cite{Dias:2015nua}, and discretize each of our patches on a $N\times N$ Chebyshev-Gauss-Lobatto grid using transfinite interpolation and pseudospectral collocation, for a total grid size of $(N+N+N+N+N)\times N$. For $N=60$, the associated grid points are explicitly displayed in Fig.~\ref{fig:map}, using a different colour code for each patch.  
After discretization, the Newton-Raphson equation and boundary conditions are solved with LU decomposition, using the $\beta=0$ (static) solution of \cite{Dias:2023rde} as a first starting seed.

\section{Further results}
\begin{figure}
    \centering
     \includegraphics[width=0.43\textwidth]{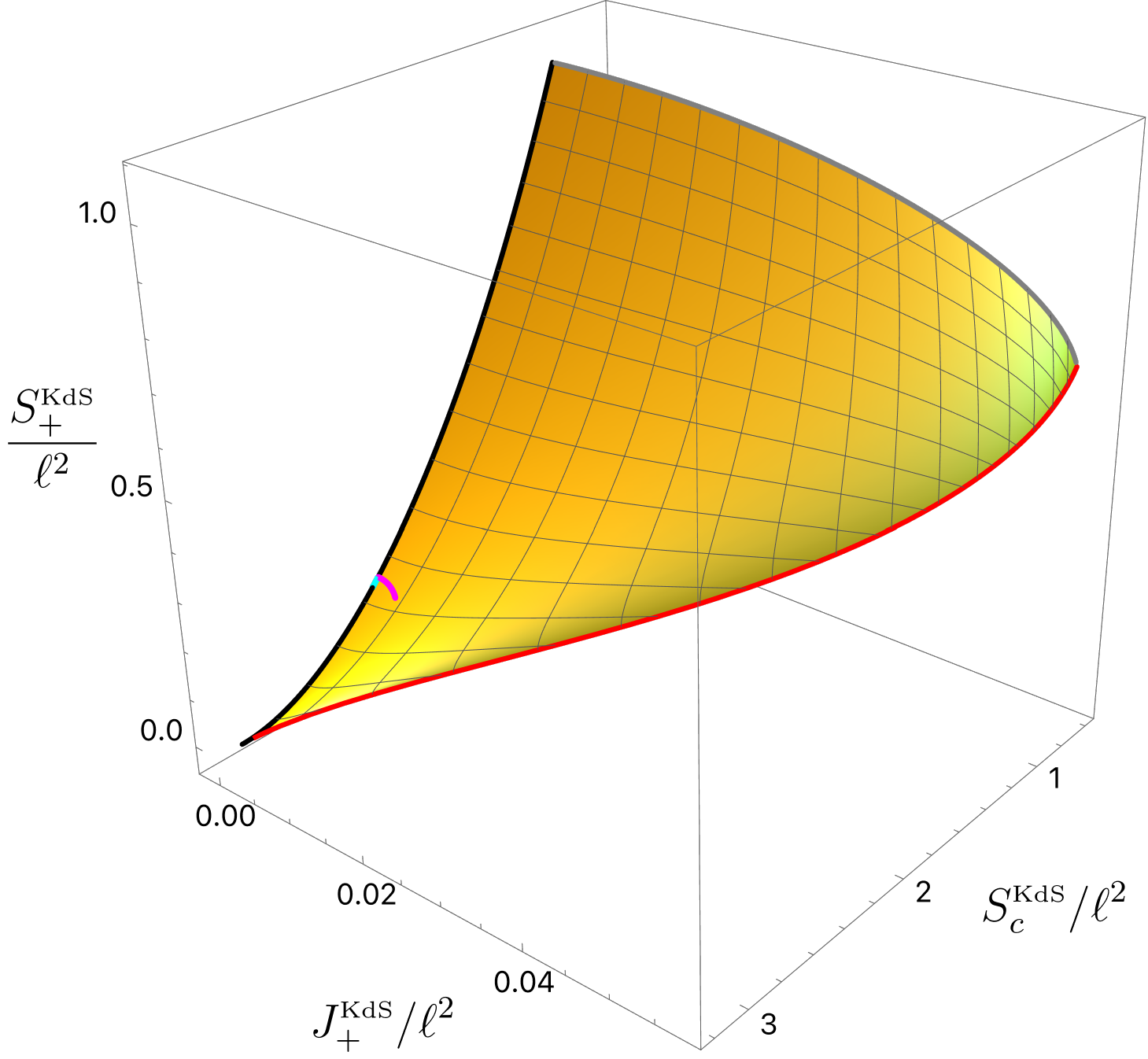}\hspace{1.5cm}
    \includegraphics[width=0.42\textwidth]{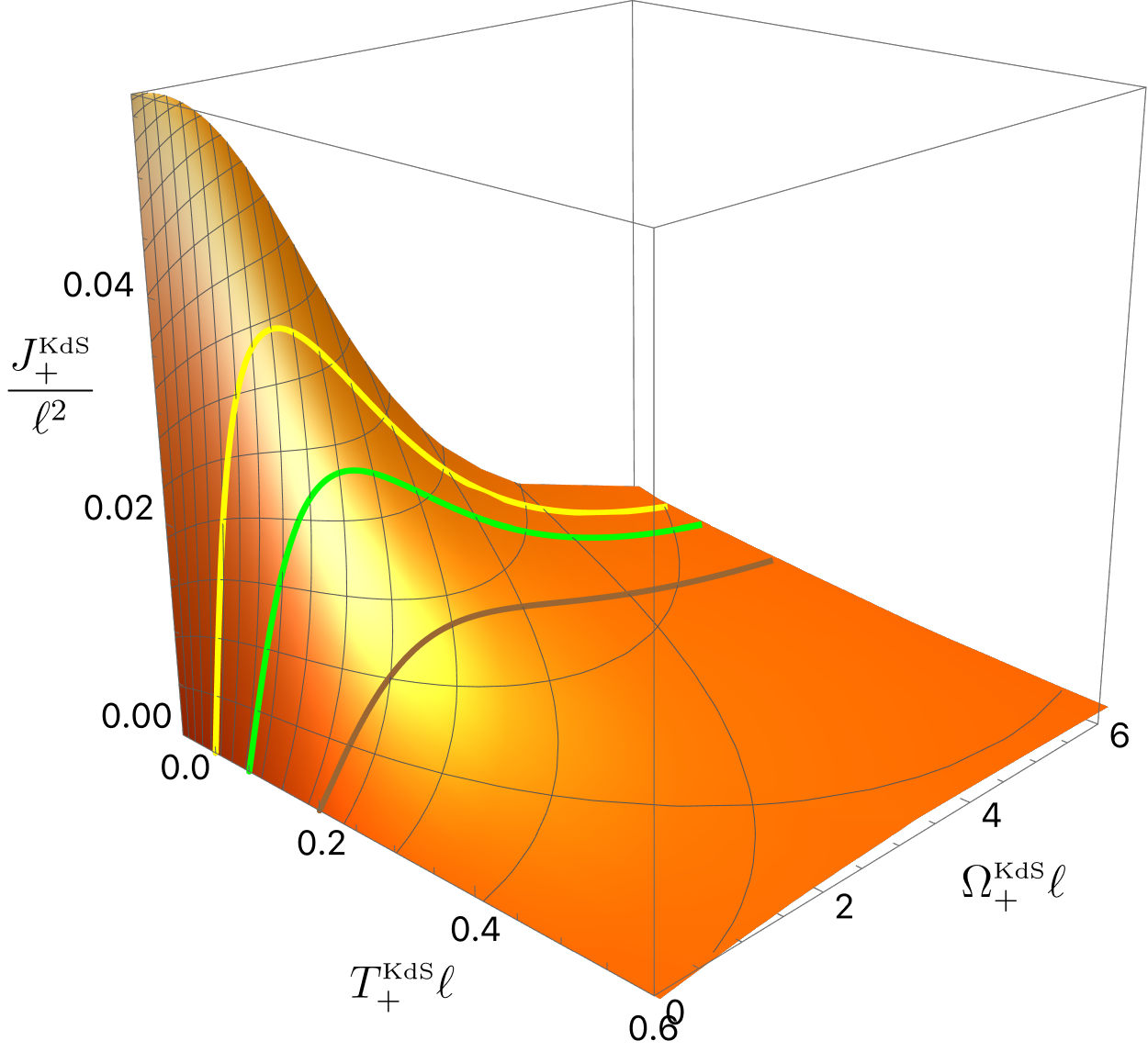}
    \caption{Thermodynamic properties of Kerr-de Sitter.}
    \label{fig:Kerr-dS}
\end{figure}

Let us review the Kerr de Sitter black hole in order to make comparisons with the spinning binaries.  We begin by introducing the parameters $y_+=\frac{r_+}{r_c}$ and $\tilde{a}=\frac{a}{r_c}$, where $r_+$ and $r_c$ are the horizon radius and cosmological horizon radius, and $a$ is the rotation parameter.  Then in a frame with a non-rotating cosmological horizon, the temperature, entropy, angular velocity, and angular momenta of the event ($+$) and cosmological ($c$) horizons of the Kerr-dS (KdS) black hole are:
\begin{align}\label{ThermoKerrdS}
& T_+^{\hbox{\tiny KdS}}\ell = \frac{(1-y_+) \left(2 y_+^3+\left(1-3 \tilde{a} ^2\right) y_+^2-2 \tilde{a} ^2 y_+-\tilde{a} ^2 \left(1+\tilde{a}^2\right)\right)}{4 \pi  \sqrt{y_+} \sqrt{y_+-\tilde{a} ^2} \left(y_+^2+\tilde{a}^2\right) \sqrt{y_+^2+y_++1+\tilde{a}^2}}
 \,, \qquad  \frac{S_+^{\hbox{\tiny KdS}}}{\ell^2} =  \frac{\pi  \left(y_+-\tilde{a} ^2\right) \left(y_+^2+\tilde{a}^2\right)}{y_+^3+y_+^2+y_++2 \tilde{a} ^2 y_+-\tilde{a} ^4} \,, \nonumber\\
&   \Omega_+^{\hbox{\tiny KdS}}\ell =  \frac{\tilde{a}  \left(1-y_+^2\right) \left(y_+^3+y_+^2+y_++2 \tilde{a} ^2 y_+-\tilde{a} ^4\right)}{\left(1+\tilde{a}^2\right) \sqrt{y_+} \sqrt{y_+-\tilde{a} ^2} \left(y_+^2+\tilde{a}^2\right) \sqrt{y_+^2+y_++1+\tilde{a}^2}}
 \,, \quad
  \frac{J_+^{\hbox{\tiny KdS}}}{\ell^2}=  \frac{\tilde{a}  \left(1+\tilde{a}^2\right) (y_++1) \left(y_+-\tilde{a} ^2\right) \left(y_+^2+\tilde{a}^2\right)}{2 \left(y_+^3+y_+^2+y_++2 \tilde{a} ^2 y_+-\tilde{a} ^4\right)^2}\,; \nonumber\\
&
T_c^{\hbox{\tiny KdS}}\ell =  \frac{(1-y_+) \left(\left(1-\tilde{a} ^2\right) y_+^2+2 \left(1-\tilde{a} ^2\right) y_+-\tilde{a} ^2 \left(\tilde{a} ^2+3\right) \right)}{4 \pi  \left(1+\tilde{a}^2\right) \sqrt{y_+} \sqrt{y_+-\tilde{a} ^2} \sqrt{y_+^2+y_++1+\tilde{a}^2}}
 \,, \qquad
 \frac{S_c^{\hbox{\tiny KdS}}}{\ell^2} =  \frac{\pi  \left(1+\tilde{a}^2\right) \left(y_+-\tilde{a} ^2\right)}{y_+^3+y_+^2+y_++2 \tilde{a} ^2 y_+-\tilde{a} ^4}\,, \nonumber\\
&   \Omega_c^{\hbox{\tiny KdS}} \ell = 0 \,, \qquad   \frac{J_c^{\hbox{\tiny KdS}}}{\ell^2}=  \frac{J_+^{\hbox{\tiny KdS}}}{\ell^2} \,.
\end{align}
Regular Kerr-dS black holes exist for $0<y_+\leq 1$ and $0 \leq \tilde{a}\leq \tilde{a}_{\rm ext}(y_+)$ where $\tilde{a}_{\rm ext}$ is the value of $\tilde{a}$ at extremality where  $T_+^{\hbox{\tiny KdS}}=0$.
In the left panel of Fig.~\ref{fig:Kerr-dS}, we display the microcanonical phase diagram with the Kerr-dS black hole family. This plot displays the full Kerr-dS family of solutions. The black line ($J_+^{\hbox{\tiny KdS}}=0$) describes the Schwarzschild-dS solution, the red line describes the extremal Kerr-dS family ($T_+^{\hbox{\tiny KdS}}=0$) and the grey curve describes  Kerr-dS black holes in the Nariai limit $y_+=1$ (\emph{i.e.} $r_+=r_c$).

To produce the microcanonical phase diagram displayed in Fig.~\ref{fig:micro} of the main text, we had to find the corresponding Kerr-dS (or Schwarzschild-dS) solution with the same $S_c/\ell^2$ and same angular momentum $J/\ell^2$ as the spinning binary. For that, we used \eqref{ThermoKerrdS} to find the matching Kerr-dS parameters $(y_+,\tilde{a})$, and then used these to find the event horizon entropy of this Kerr-dS black hole. This generates the magenta curve in Fig.~\ref{fig:micro} of the main text or the cyan and magenta curves in the left panel of Fig.~\ref{fig:micro} (note that for the cyan curve one has $a=J_+^{\hbox{\tiny KdS}}=0$, \emph{i.e.\!} it describes  Schwarzschild-dS black holes) that are also displayed in the left panel of Fig.~\ref{fig:Kerr-dS} (the cyan curve on top of the black curve is barely seen since it has very small length). As it is clear from Fig.~\ref{fig:micro} in the main text, the 2-parameter families of regular  (anti-)aligned spinning binaries are described by surfaces that are well below the Kerr-dS surface of Fig.~\ref{fig:Kerr-dS} (for values of $J/\ell^2$ and $S_c/\ell^2$ where they co-exist).

In the main text, we pointed out that for Kerr-dS black holes with fixed temperature, the Komar angular momentum $J_+^{\hbox{\tiny KdS}}/\ell^2$ first increases with $\Omega_+^{\hbox{\tiny KdS}}\ell$ till it reaches a maximum, and then decreases as $\Omega_+\ell$ increases further. This is explicitly shown in the right panel of Fig.~\ref{fig:Kerr-dS} where we plot $J_+^{\hbox{\tiny KdS}}/\ell^2$ as a function of $T_+^{\hbox{\tiny KdS}}\ell$ and $\Omega_+^{\hbox{\tiny KdS}}\ell$ and display three families with constant  $T_+^{\hbox{\tiny KdS}}\ell=0.05,0.1,0.2$ (yellow, green, brown curves); note that Ker-dS exists for higher values of temperature and angular velocity than those shown.

\begin{figure}[ht]
    \centering
    \includegraphics[width=0.435\textwidth]{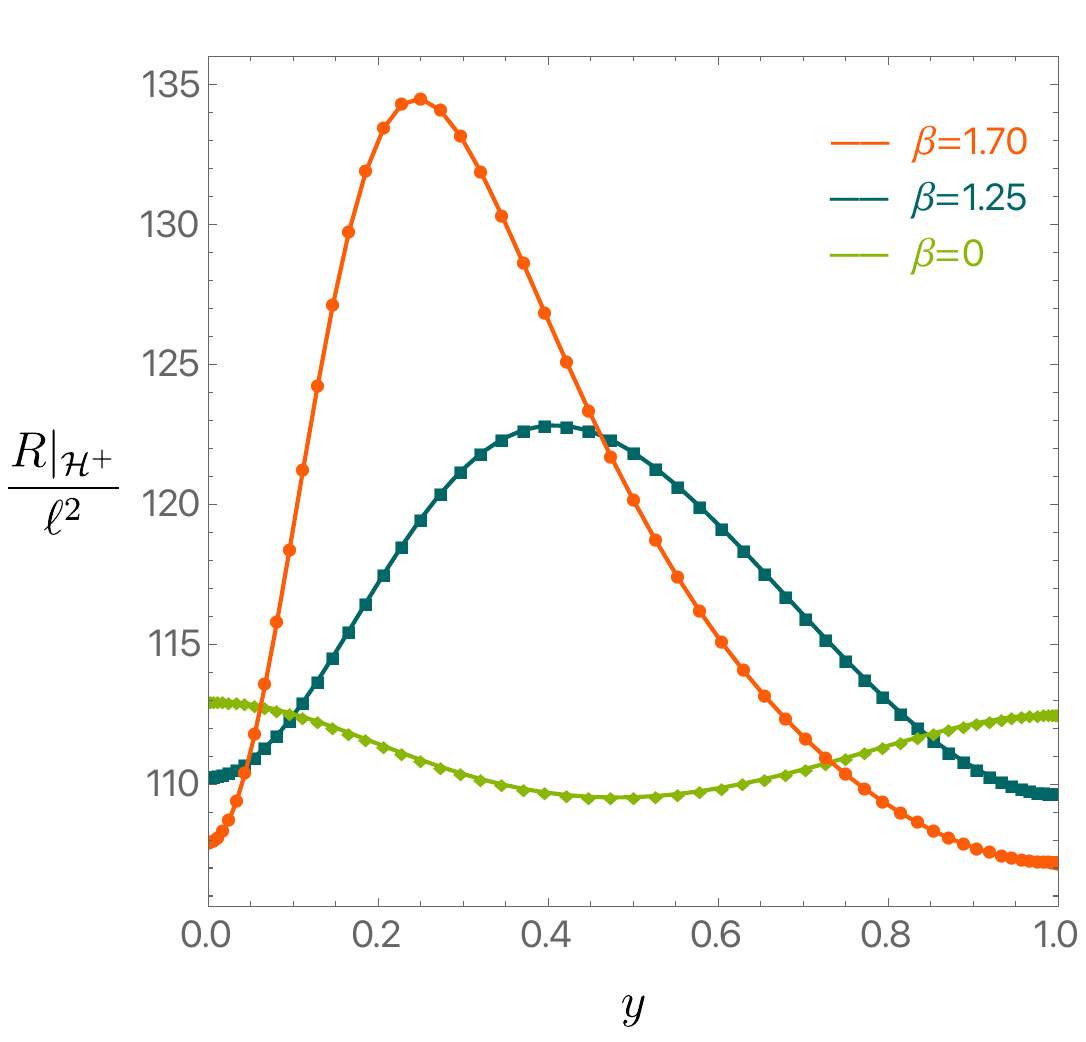} \hspace{1.5cm}
     \includegraphics[width=0.42\textwidth]{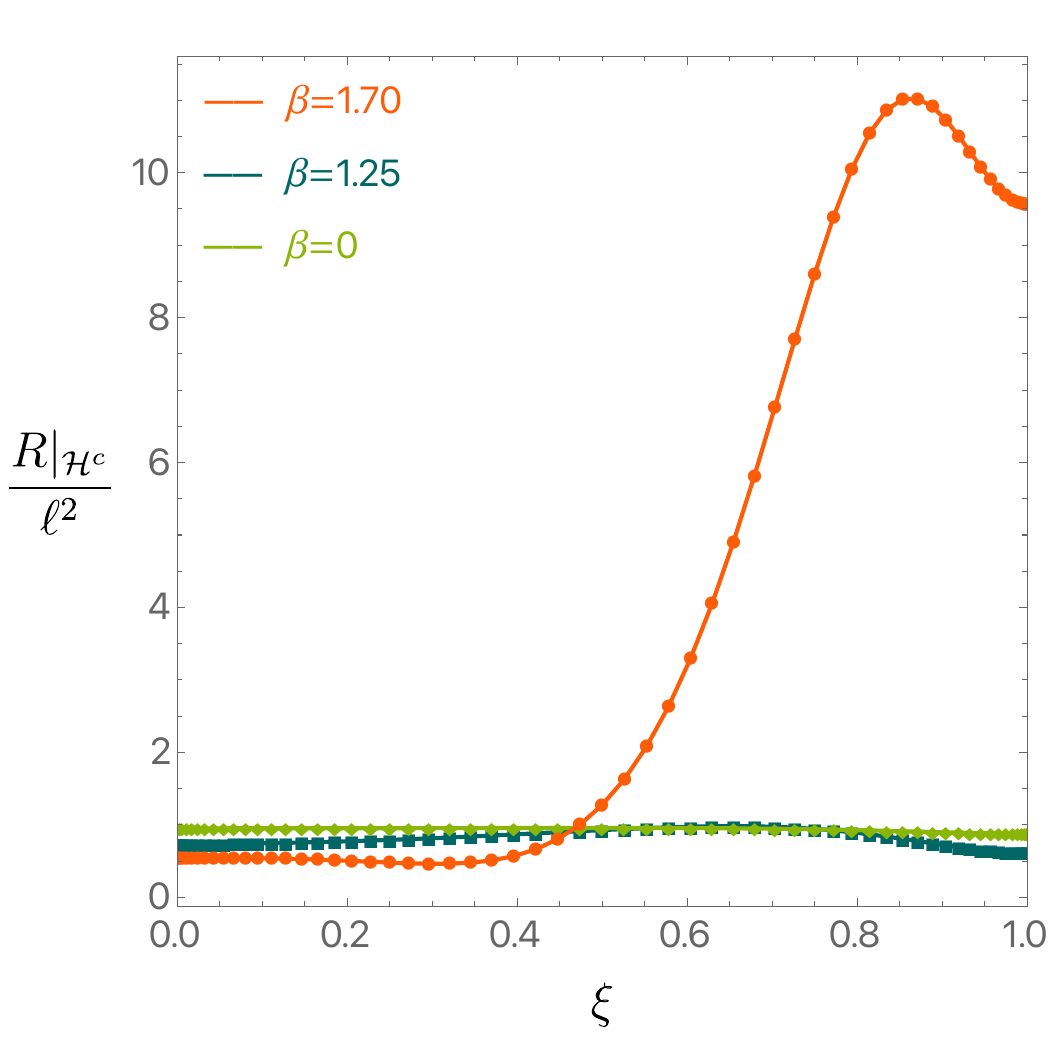}
    \caption{Ricci curvature evaluated at the black hole horizon (left panel) and at the cosmological horizon (right panel)}
    \label{fig:RicciHorizons}
\end{figure}

\begin{figure}[b]
    \centering
    \includegraphics[width=0.43\textwidth]{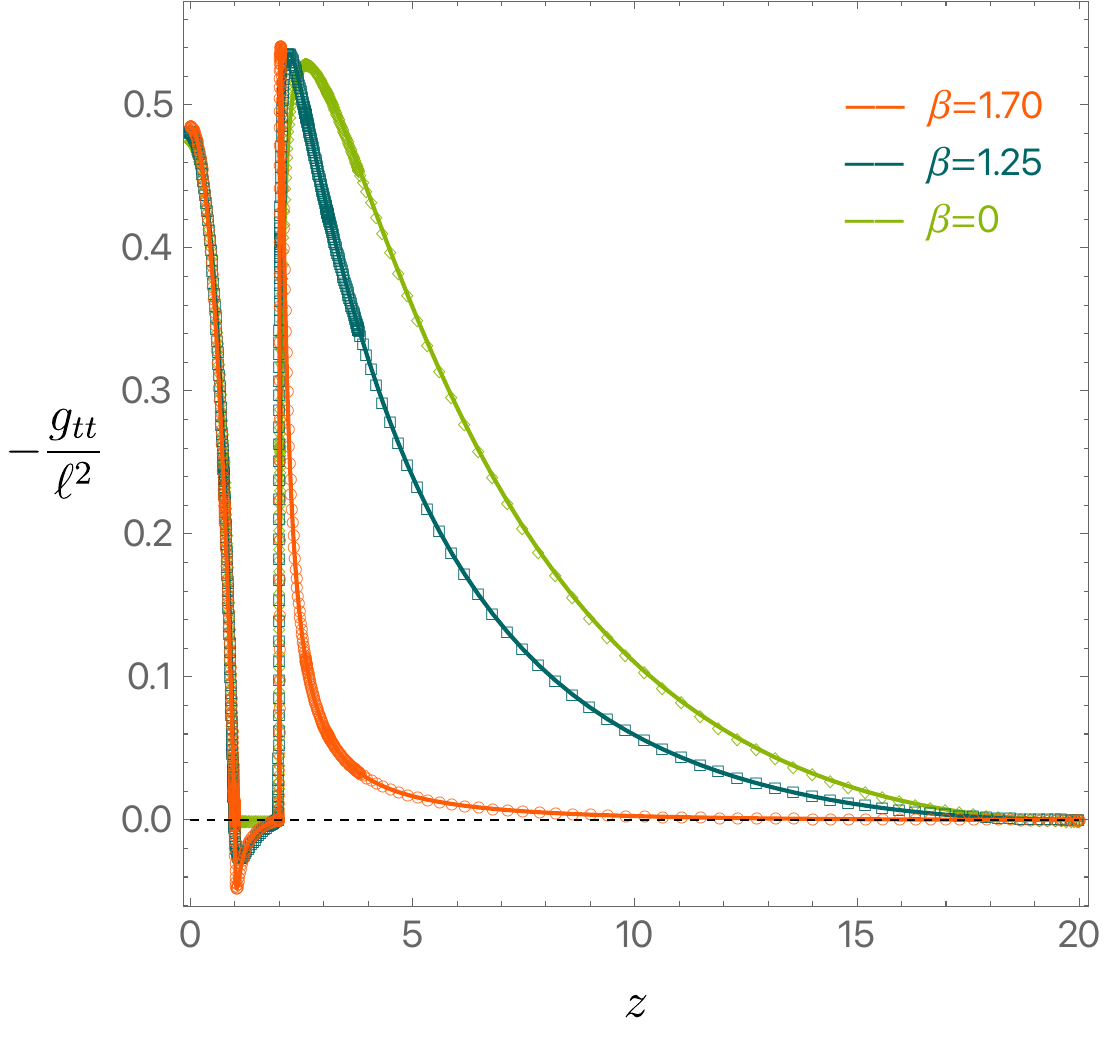} \hspace{1.5cm}
     \includegraphics[width=0.42\textwidth]{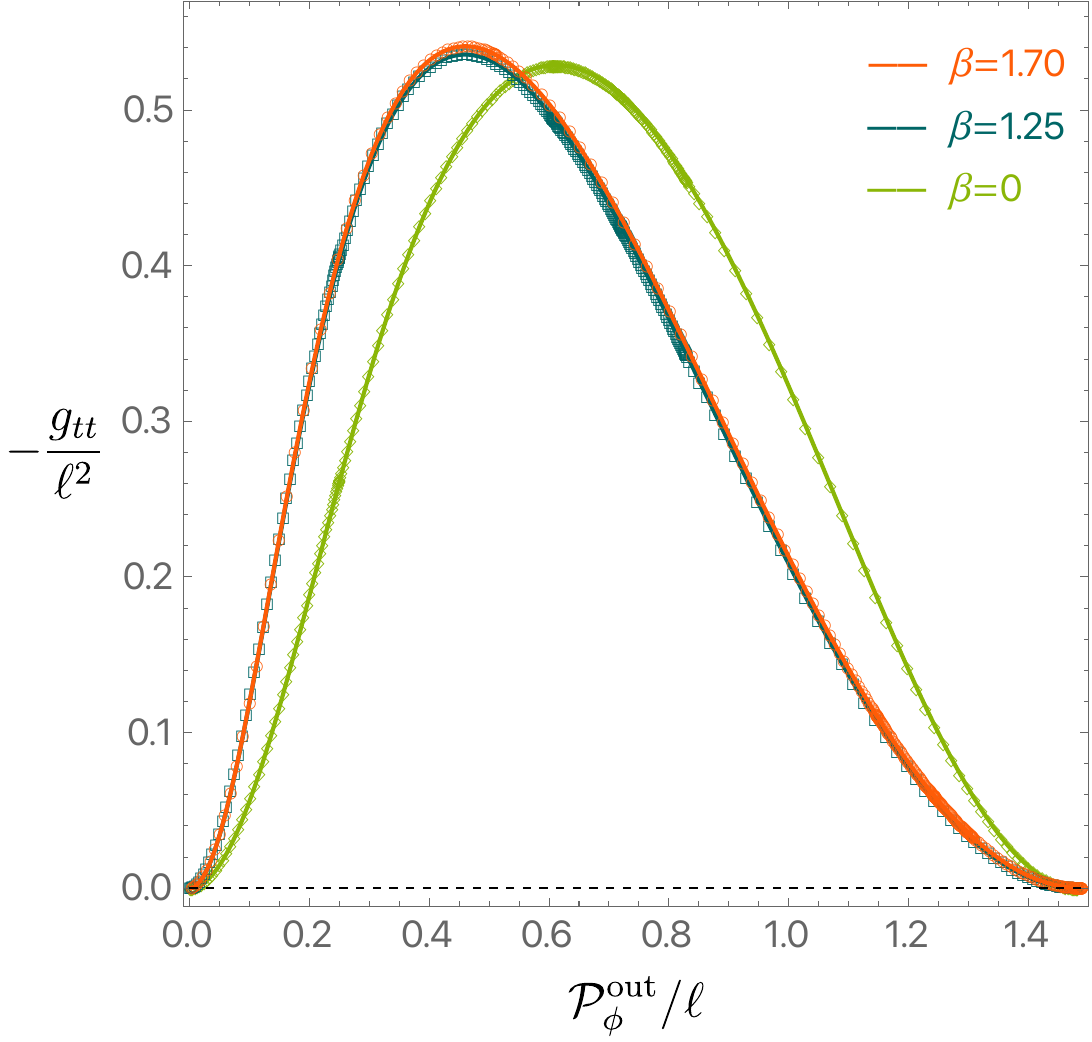}
    \caption{Left panel: Gauge invariant metric component $-g_{tt}$, evaluated along the $z-$axis containing the inner, black hole horizon and outer axis. Right panel: $-g_{tt}$ as a function of the proper distance $\mathcal{P}_{\phi}^{\rm out}$ between the event horizon and a point outside it.}
    \label{fig:gtt}
\end{figure}

We have designed our numerical code such that the binary solutions are parametrized by the black hole temperature and angular velocity. In particular, some of the solutions we present have constant temperature $T_+\ell=\frac{3.75}{2\pi}$ and $|\Omega_+^{(i)}\ell|=\beta$ with $\beta\in [0,1.72]$. We stopped at $\beta=1.72$ because it becomes increasingly more difficult to find numerical solutions with higher $\beta$. This occurs because some of the functions start developing very large gradients and/or a hierarchy of scales between different regions of the integration domain. However, do not find no evidence for the appearance of curvature singularities beyond $\beta>1.72$.  One such curvature invariant that we monitored is the pullback of the Ricci curvature into the black hole ($x=1$) and cosmological horizons ($\rho=2/\lambda$), which are shown in the left and right panels of Fig.~\ref{fig:RicciHorizons}, for three different values of $\beta$. These quantities remain clearly finite as $\beta$ grows which suggests that our binaries should exist for even higher values of $\beta$ without developing curvature singularities.

An important gauge invariant quantity of the binary system is the time-time component of the metric $g_{tt}$. In the left panel of Fig.~\ref{fig:gtt} we plot $-g_{tt}$ as a function of the Weyl coordinate $z$ along the axis that connects the two black holes (\emph{i.e.\!} with Weyl coordinate $r=0$) and that contains the inner $\partial_\phi$-axis ($z\in [0,1]$), black hole horizon ($z\in [1,1/k]$) and outer $\partial_\phi$-axis ($z \in [1/k,2/\lambda]$) for anti-aligned binaries with $k=0.5$ for three different values of $\Omega^{(1)}_+=\Omega^{(2)}_+=\beta$. We first notice the existence of an ergoregion with $-g_{tt}<0$. As expected, we see that this ergoregion is absent for $\beta=0$, and the  minimum of  $-g_{tt}$ becomes more negative as $\beta$ increases. One also finds that $-g_{tt}$ becomes considerably more spiked around its local maximum outside the black hole horizon (\emph{i.e.\!} just above $z=1/k=2$) and then decays faster with $z$ as $\beta$ increases. This partially justifies why it becomes increasingly more difficult to find our numerical solutions as $\beta$ grows. But it is important to note that this does not imply the development of any pathology. Indeed, if instead of the (gauge dependent) coordinate $z$, we plot $-g_{tt}$ as a function of the (gauge invariant) proper length $\mathcal{P}_{\phi}^{\rm out}=\int_{1/k}^{z}\sqrt{g_{\phi\phi}|_{r=0}}\mathrm{d}\tilde{z}$ along the outer $\partial_\phi$-axis, as we do in the right panel of Fig.~\ref{fig:gtt},  one finds that the qualitative behaviour of the curve (in particular, its width) does not change significantly as $\beta$ increases. Anti-aligned spinning binaries  with $\Omega^{(1)}_+=-\Omega^{(2)}_+=\beta$ have similar qualitative behaviour.

\begin{figure}[b]
\centering
\includegraphics[width=0.44\textwidth]{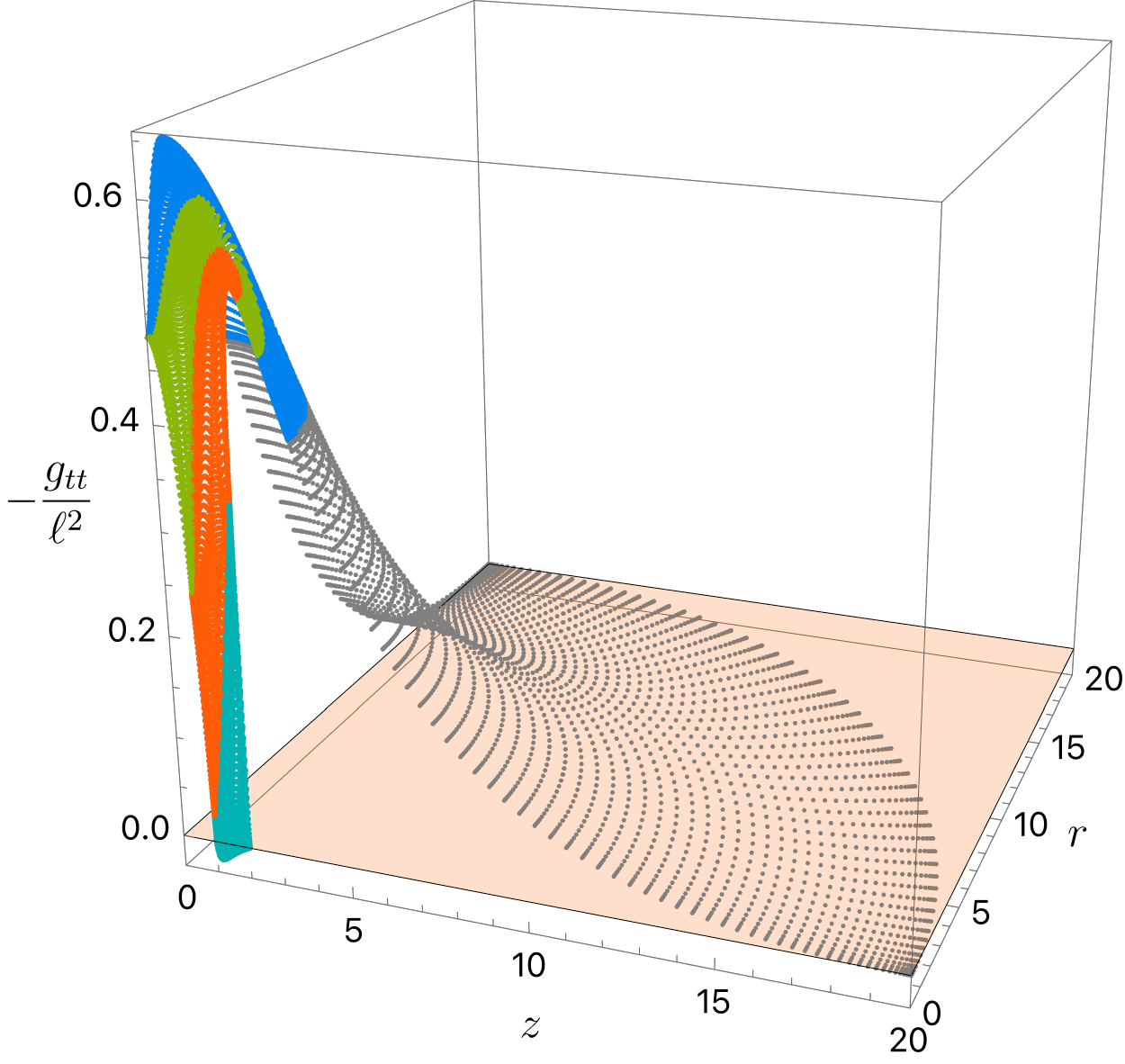}\hspace{1cm}
\includegraphics[width=0.47\textwidth]{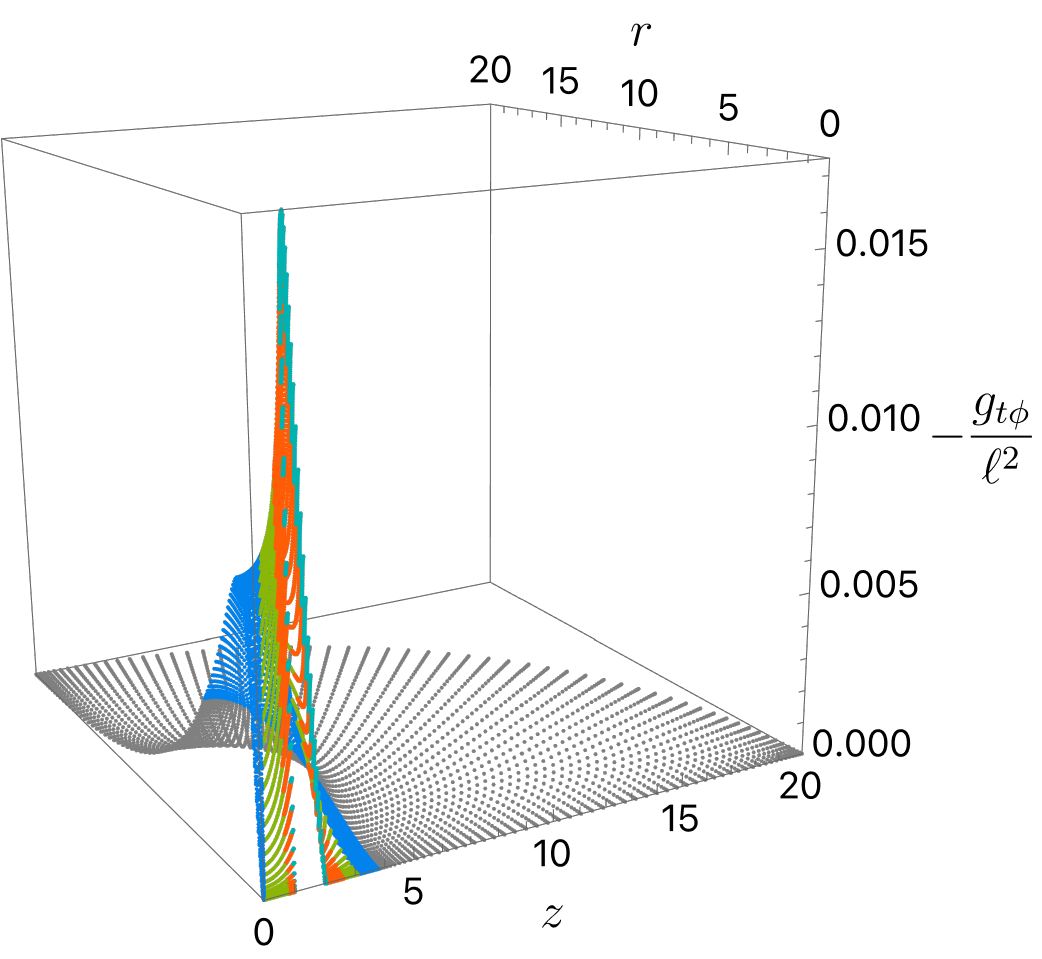}
\caption{The gauge invariant metric functions $-g_{tt}$ and $g_{\phi\phi}$ for $T_+/T_c=3.75$ ($\lambda=0.1, k=0.5$) and $\Omega_+^{(1)}\ell=\Omega_+^{(2)}\ell\equiv \beta=1.0$.}\label{fig:functionsk05b1}
\end{figure}

\begin{figure}[t]
\centering
\includegraphics[width=0.44\textwidth]{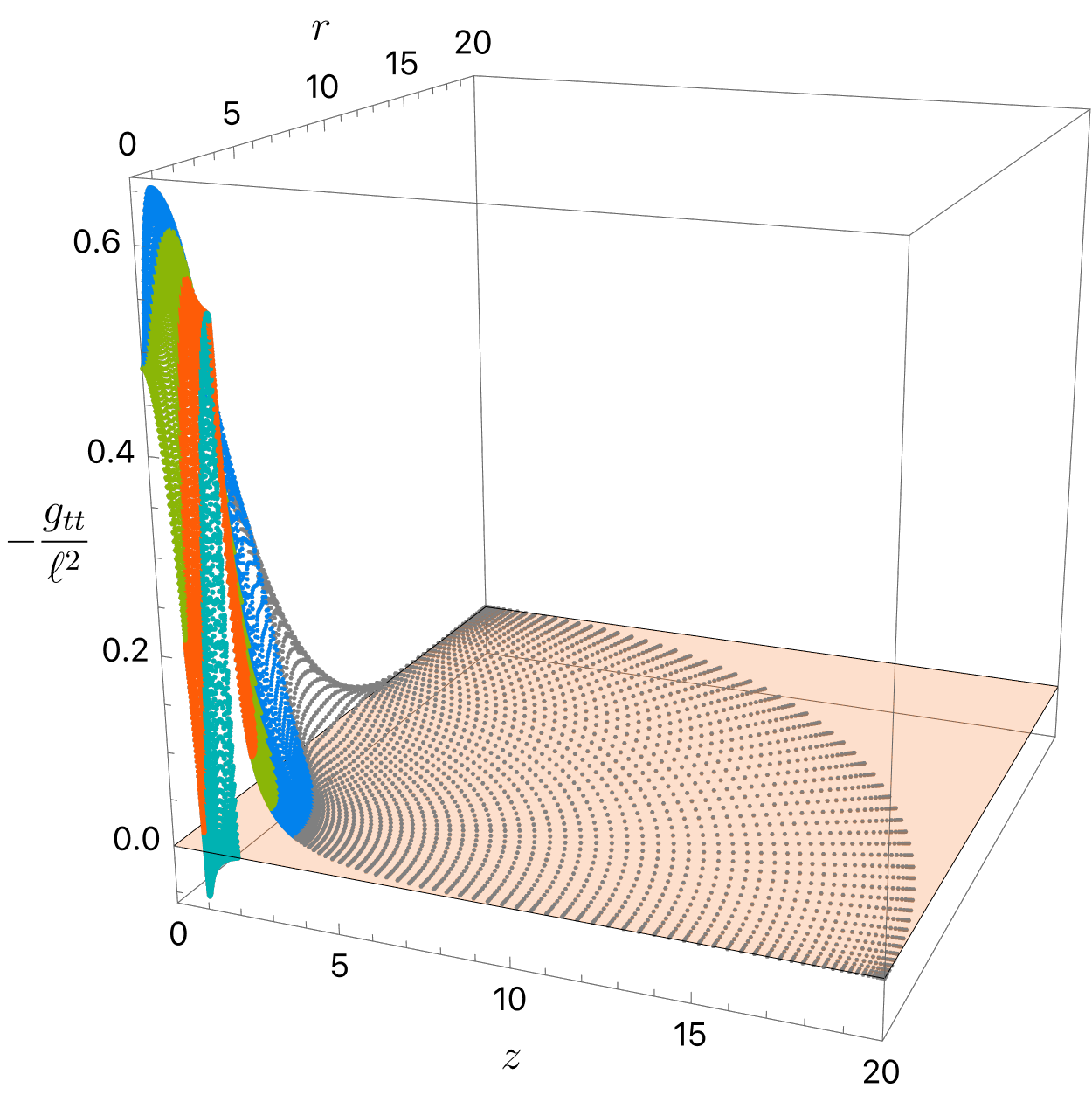}\hspace{1cm}
\includegraphics[width=0.47\textwidth]{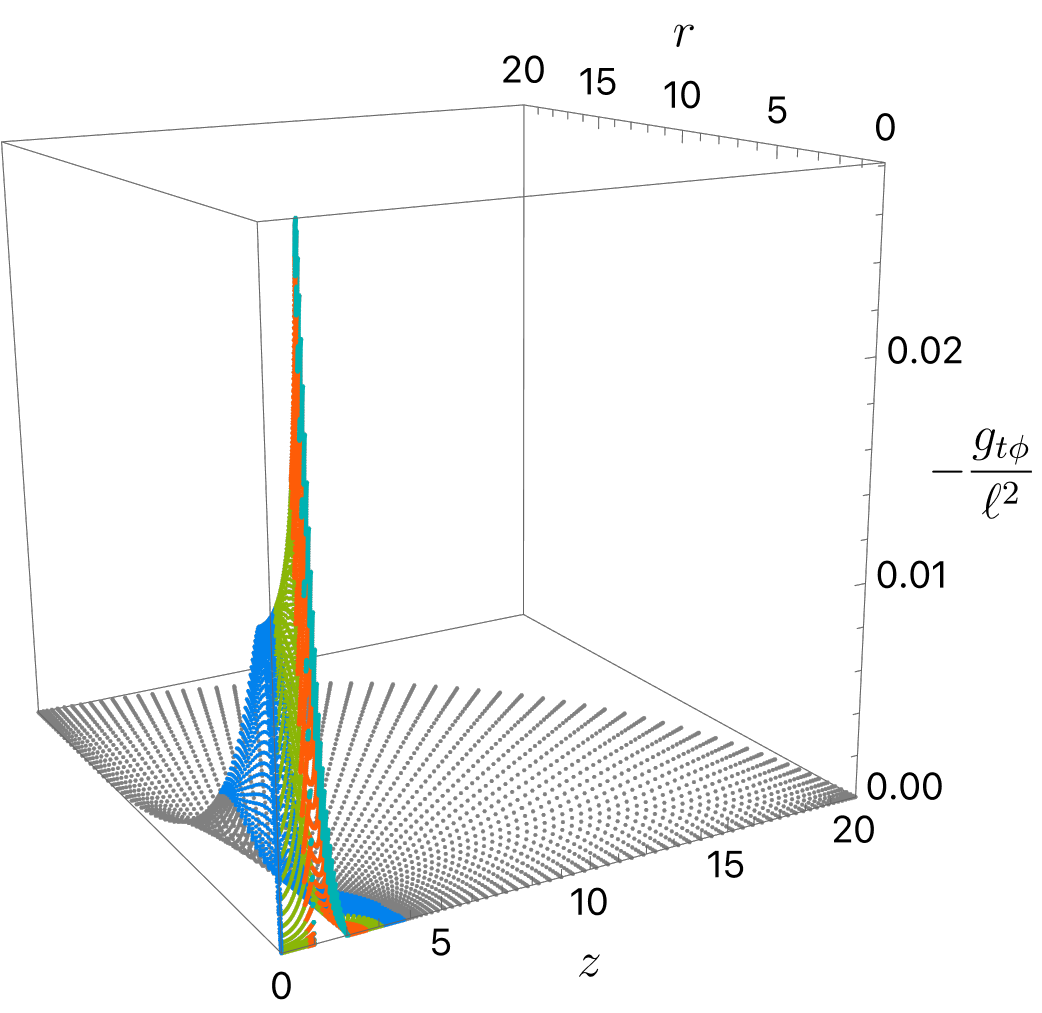}
\caption{The gauge invariant metric functions $-g_{tt}$ and $-g_{t\phi}$ for $T_+/T_c=3.75$ ($\lambda=0.1, k=0.5$) and $\Omega_+^{(1)}\ell=\Omega_+^{(2)}\ell\equiv \beta=1.7$.}\label{fig:functionsk05b1p7}
\end{figure}

To have a more holistic illustration of how our solutions look like, (besides Fig.~\ref{fig:level} in the main text) in Figs.~\ref{fig:functionsk05b1}$-$\ref{fig:functionsk05b1p7} we display the two metric functions $-g_{tt}$ and $-g_{t\phi}$  (that are gauge invariant since $\partial_t$ and $\partial_\phi$ are Killing vector fields)  as a function of the original cylindrical Weyl coordinates $(r,z)$ for two representative aligned spinning binaries with black hole temperature $T_+/T_c=3.75$ and angular velocity $\Omega_+^{(1)}\ell=\Omega_+^{(2)}\ell=1.0$ (Fig.~\ref{fig:functionsk05b1}) and $\Omega_+^{(1)}\ell=\Omega_+^{(2)}\ell=1.7$ (Fig.~\ref{fig:functionsk05b1p7}). The system is $\mathbb{Z}_2$ symmetric about $z=0$, \emph{i.e.\!} $g_{ab}(r,-z)=\pm g_{ab}(r,z)$ (the minus sign holds only for the component $ab=t\phi$ in the anti-aligned binary) and thus we just display the solution for $z\geq 0$. The five different colours represent the five distinct patches that we use and the colour code is the same one displayed in Fig.~\ref{fig:map}. Note that, as required, one always has a smooth transition between patches. Anti-aligned spinning binaries  with $\Omega^{(1)}_+=-\Omega^{(2)}_+=\beta$ have similar qualitative behaviour.


Recall that in this cylindrical Weyl coordinate chart $(r,z)$, we have a ``rod structure" where the rotation axis and the black hole horizons are all located at $r=0$. More concretely, at $r=0$, the black horizon lies in the region $z\in (1,1/k)$ and thus, the norm of the horizon generator $K=\partial_t+\Omega_+^{(i)}\partial_\phi$ vanishes here (not plotted). The ergoregion (where $-g_{tt}<0$) is also along this segment $z\in (1,1/k)$  at $r=0$ but extends for $r>0$ in a neighbourhood of this rod. Figs.~\ref{fig:functionsk05b1}$-$\ref{fig:functionsk05b1p7} indeed show that $-g_{tt}$ is negative in these ergoregions and that the negative absolute minimum of $-g_{tt}$ (that is at $r=0$ as better pinpointed by the magenta inverted triangles in Fig.~\ref{fig:level}) increases as $\beta$ increases. Moreover, still at $r=0$, the inner segment of the axis between the black holes is in the region $z\in(-1,1)$ and the outer segments of the axis are in $z\in(1/k,\infty)$ and $z\in(-\infty,-1/k)$: not shown (see \cite{Dias:2023rde}), $g_{\phi\phi}$ vanishes in these segments. The solution has a cosmological horizon at $\sqrt{r^2+z^2}=2/\lambda$ (with temperature $T_c=1/(2\pi)$) where $g_{tt}$ vanishes (since this horizon is generated by  $\partial_t+\Omega_c\partial_\phi$ with $\Omega_c=0$) as clearly identified in the left panels of Figs.~\ref{fig:functionsk05b1}$-$\ref{fig:functionsk05b1p7} .

In the right panels of Figs.~\ref{fig:functionsk05b1}$-$\ref{fig:functionsk05b1p7} we plot $-g_{t\phi}$. This function vanishes everywhere when $\beta=0$. When the components of the binary are spinning, $-g_{t\phi}$ vanishes at $r=0$ along the inner $z\in(0,1)$ and outer  $z\in(1/k,\infty)$ segments of the axis. It also vanishes at the cosmological horizon $\sqrt{r^2+z^2}=2/\lambda$ since $\Omega_c=0$, but is otherwise non-zero. In particular, it is non-vanishing at $r=0$ along $z\in (1,1/k)$ inside of which it attains its maximum value, and this maximum value increases as $\beta$ increases. Note that $-g_{t\phi}$ is an even function of $z$ for aligned spinning binaries (the case that is shown in Figs.~\ref{fig:functionsk05b1}$-$\ref{fig:functionsk05b1p7}), but is an odd function of $z$ in the anti-aligned binary case (not shown).

\section{Convergence Tests}
In this section, we show that the norm $\chi\equiv \xi^a \xi_a$ of the DeTurck vector vanishes in the continuum limit, as expected for a solution of the Einstein-DeTurck equation that is not a Ricci soliton (\emph{i.e.\!} that is instead a true solution to the Einstein equation). Additionally, we find  exponential convergence, which is consistent with the use of Chebyshev pseudospectral collocation methods.

\begin{figure}[ht]
    \centering
    \includegraphics[width=0.45\textwidth]{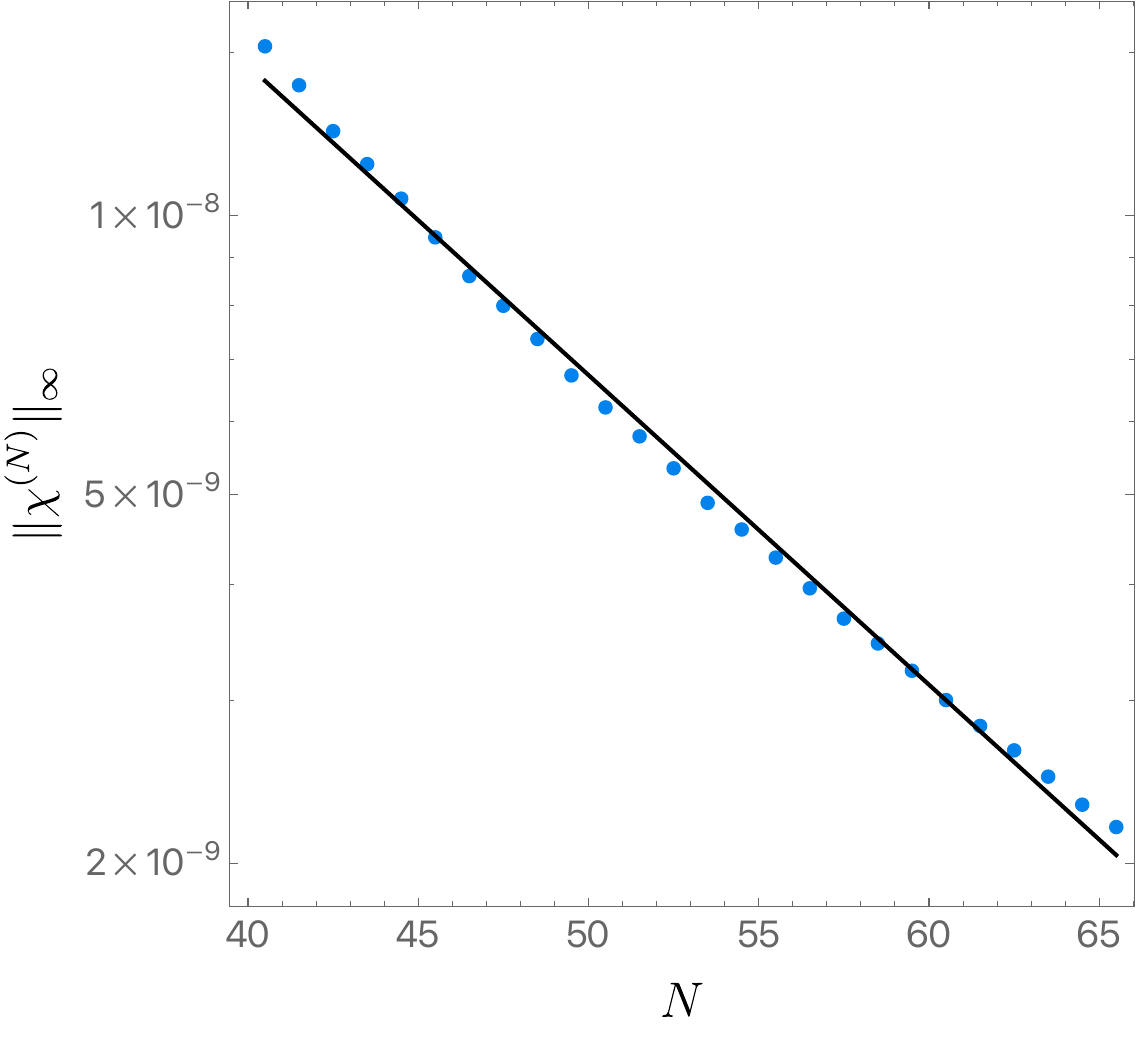}
    \caption{Convergence test showing both the exponential accuracy of our pseudospectral numerical method and the fact that we are not converging to a Ricci soliton.}
    \label{fig:convergence}
\end{figure}

Let $\chi^{(N)}$ be $\chi$ computed on a (5-patched) grid with $(N+N+N+N+N)\times N$ spectral collocation points. For concreteness, we take $\alpha$ as given in~\eqref{Spin:alphaspecial}, $\lambda=1/10$, $k=1/2$ and $\beta=1.5$. In Fig.~\ref{fig:convergence} we show $\lVert \chi^{(N)}\rVert_{\infty}$ as a function of $N$ in a $\log$-plot. The solid black line shows the best $\chi^2$-fit to a straight line in the $\log$-plot, and yields
\begin{equation}
f(N)=-14.97022 - 0.07694\,N\,.
\end{equation}
The exponential trend is clear and confirms that the Einstein-DeTurck solution is converging to a true solution of the Einstein equation (and not to a Ricci soliton with finite $\chi$).

\bibliographystyle{utphys-modified}
\bibliography{papersspinning}
\end{document}